%% file: nedge.tex
\RequirePackage[left,modulo]{lineno}

\documentclass[aps,twocolumn,showpacs,amsmath,prd,amssymb,superscriptaddress,nofootinbib]{revtex4}

\modulolinenumbers[2]

\pdfoutput=1

\usepackage{graphicx}
\usepackage{dcolumn}
\usepackage{bm}

\def\cpkkd{\rm{kg^{-1}keV_{ee}^{-1}day^{-1}}}
\def\mwimp{\rm{m_{\chi}}}

\def\munu{\mu_{\nu}}
\def\mub{\mu_{\rm B}}
\def\sasix{\rm{SA_{6}}}
\def\sa12{\rm{SA_{12}}}
\def\trigeff{\varepsilon _{\rm Trig}}
\def\daqeff{\varepsilon _{\rm DAQ}}
\def\CReff{\varepsilon _{\rm CR}}
\def\ACeff{\varepsilon _{\rm AC}}
\def\bfeff{\varepsilon _{\rm bf}}

\def\nuebar{\bar{\nu}_e}
\def\amprms{\sigma_{A}}
\def\qrms{\sigma_{Q}}
\def\eamp{A_{max}}
\def\echarge{Q_{area}}

\def\keVee{\rm keV_{ee}}
\def\eVee{\rm eV_{ee}}

\def\effbs{\varepsilon _{\rm BS}}
\def\lmbdbs{\lambda _{\rm BS}}
\def\b0{\rm B_0}
\def\s0{\rm S_0}

\begin{document}

\title{
Characterization and Performance of
Germanium Detectors with sub-keV Sensitivities 
for Neutrino and Dark Matter Experiments
}

%

\input{author_nedge.tex}

\collaboration{TEXONO Collaboration}



\date{\today}

\begin{abstract}

Germanium ionization detectors with  sensitivities 
as low as 100~$\eVee$ (electron-equivalent energy) open new
windows for studies on neutrino and 
dark matter physics. 
The relevant physics subjects are summarized. 
The detectors have to measure physics signals 
whose amplitude is comparable to that of pedestal 
electronic noise. 
To fully exploit this new detector technique,  
various experimental issues
including quenching factors, energy reconstruction and calibration, 
signal triggering and selection as well as evaluation of
their associated efficiencies have to be attended.
The efforts and results of a research program to address 
these challenges are presented.

\end{abstract}


\maketitle

\section{Introduction}

Sensitivities on several important 
research programs in neutrino and dark matter physics 
can be significantly enhanced when
the ``physics threshold'' can be lowered
to extend the dynamic range of 
signal detection~\cite{texonoprogram,cogentppcge}.
This motivates efforts to characterize detector behavior 
and to devise analysis methods in domains 
where the amplitude of physics signals 
is comparable to that caused by fluctuations of
pedestal electronic noise.

In this article, we report on our research program and results 
on using advanced germanium (Ge) ionization detectors
to address the above mentioned issues. 
Following a survey on physics topics relevant to 
low-background and low-threshold techniques,
crucial aspects of detector operation 
and optimizations near electronic ``noise-edge'' 
are discussed.
These include studies on energy estimators and calibration, 
trigger and data acquisition rates, signal event selections and 
evaluation of their efficiencies. 

Data taken with point-contact Ge detectors 
with sub-keV sensitivities were adopted 
to establish the results. 
However, the devised techniques would also be 
applicable to other detector systems, 
and at other energy ranges. 
Unless otherwise stated,
electron-equivalent energy ($\eVee$) is used
throughout this article to denote detector response
to a measurable energy T.
The raw kinetic energy due to nuclear recoils 
is denoted by $\rm{keV_{\rm nr}}$.

Results on the characterization
and performance of Ge detectors 
are original work.
Surface background~\cite{bsel2014} and 
quenching factor~\cite{qfge} of Ge detectors 
have been discussed in the literature.
They are summarized in 
Sections~\ref{sect::bsppcge} and \ref{sect::qf}, 
respectively, for completeness and coherence. 

\begin{figure}
{\bf (a)}\\
\includegraphics[width=8.5cm]{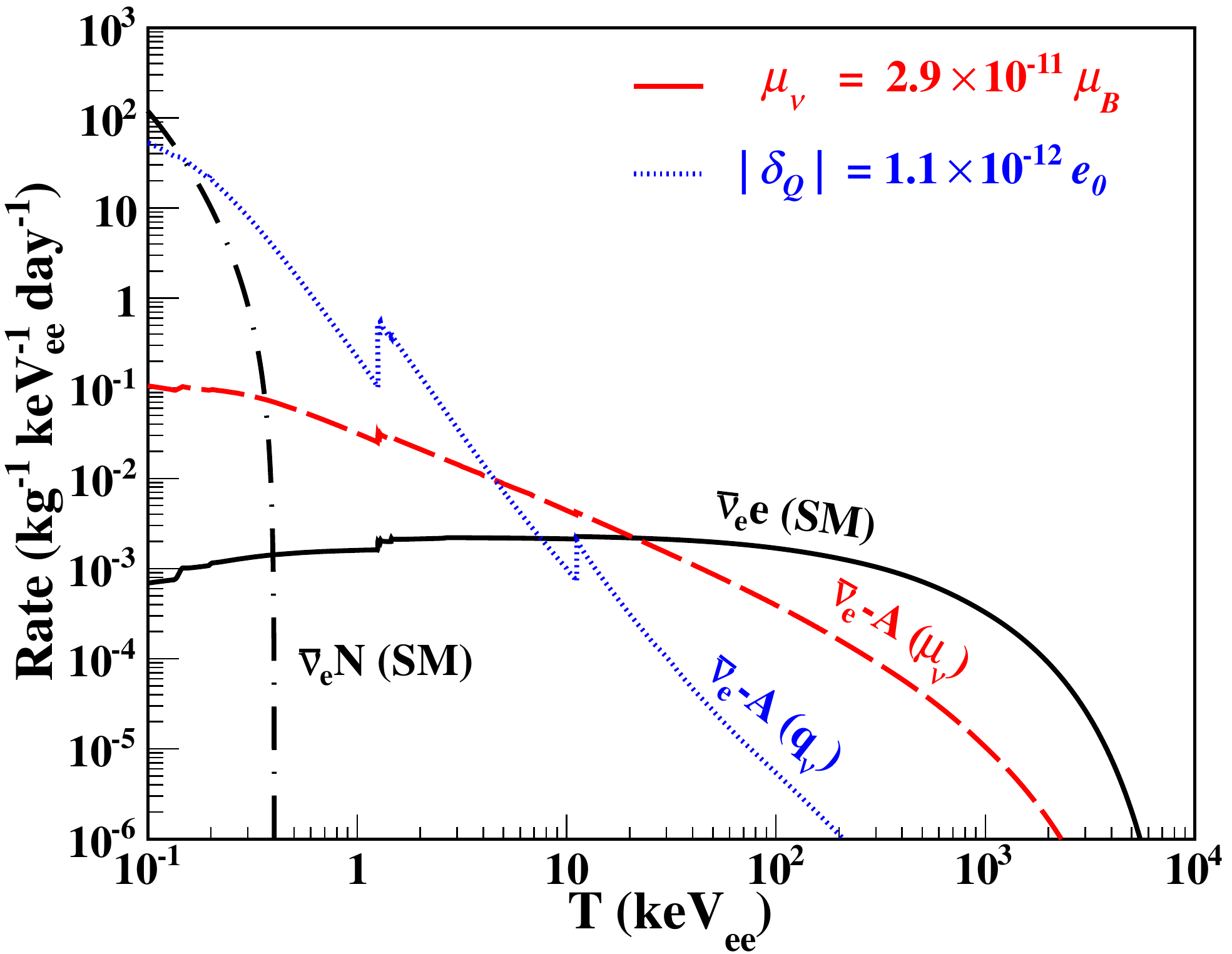}\\
{\bf (b)}\\
\includegraphics[width=8.5cm]{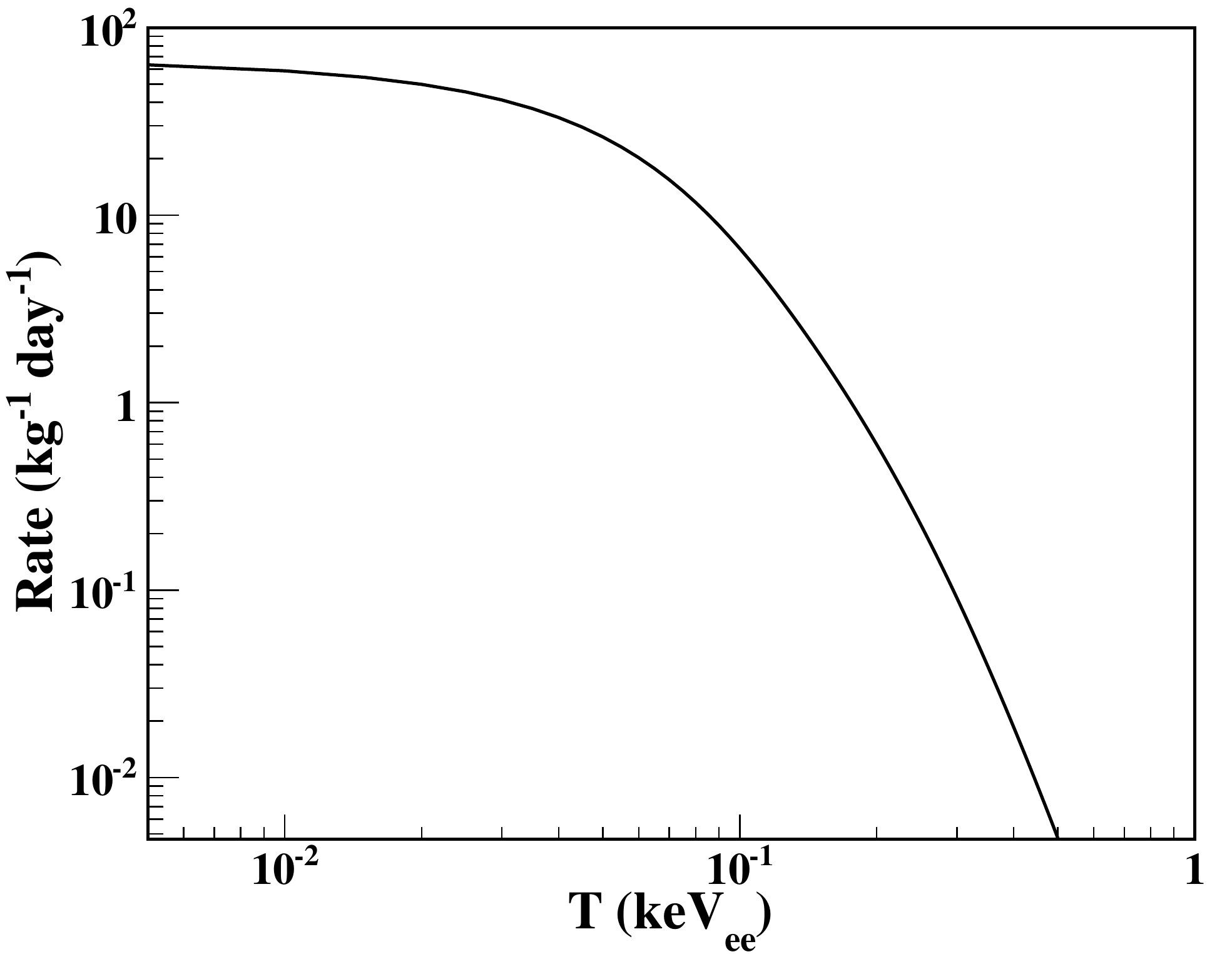}
\caption{ 
(a)
Observable spectra due to 
reactor-$\nuebar$ interactions on Ge target
with $\phi ( \nuebar ) = 10^{13}~{\rm cm^{-2} s^{-1}}$,
neutrino magnetic moment 
and neutrino milli-charge fraction 
at the current bounds from direct experimental searches:
$\munu = 2.9 \times 10^{-11} ~ \mub$ and  
$| \delta_{\rm Q} | = 1.1 \times 10^{-12}$, respectively.
Superimposed are SM $\nuebar$-e and coherent scattering $\nuebar$-N.
Quenching effects of nuclear recoils are taken into account.
(b)
Expected integral $\nuebar$-N coherent scattering 
rates due to SM contributions at the same flux, 
as a function of physics threshold, assuming realistic
detector resolution.
}
\label{fig::dNdT}
\end{figure}

\section{Scientific Motivations}
\label{sect::phys}

The objective of our research program
is to develop detectors with
modular mass of $\mathcal{O}$(1~kg), 
physics threshold 
of $\mathcal{O} ( 100 ~ \eVee )$ and background
level at threshold of $\mathcal{O}( 1~\cpkkd)$~\cite{texonoprogram}.
Germanium semiconductors in ionization mode were selected as the
detection technique.
When these ``benchmark'' specifications are fulfilled,
several important topics discussed in subsequent sections
can be experimentally pursued.

\subsection{Neutrino Electromagnetic Properties}

Investigations of 
neutrino properties and interactions
can reveal physics within 
and beyond the Standard Model (SM).
An avenue is the study
of possible neutrino electromagnetic
interactions~\cite{nuem-review}.

The neutrino magnetic moment ($\munu$) is an 
intrinsic neutrino property
that describes possible
neutrino-photon couplings via its spin~\cite{munureview,texonomunu}.
The helicity is flipped in these $\munu$-induced interactions.
Observations of $\munu$ at levels relevant to present or future
generations of experiments would strongly favor the case of 
Majorana neutrinos~\cite{naturalnessth}.
Most experimental searches of $\munu$ make use of
neutrino interactions with free electrons. 
The differential cross-section has an (1/T)-dependence,
where the measurable T is due to 
recoil kinetic energy of electrons.
The expected differential spectra 
for reactor neutrinos at a flux of 
$\phi ( \nuebar ) = 10^{13}~{\rm cm^{-2} s^{-1}}$
are shown in Figure~\ref{fig::dNdT}a
(details of reactor $\nuebar$-spectra
and their derivations are described in 
Refs.~\cite{texonomunu,texononuecsi}).
Contributions from  $\munu$ 
are enhanced as T decreases, with 
necessary modifications from
the atomic binding energy effects~\cite{nmmai,nuemai}. 

In a similar spirit, studies on neutrino ``milli-charge'' 
probe possible helicity conserving QED-like interactions.
It can be parametrized as ($\delta _{\rm Q} \cdot e_0$) 
where $\delta _{\rm Q}$
is the charge fraction and
$e_0$ is the standard electron charge.
Finiteness of $\delta _{\rm Q}$ would imply 
that neutrinos are Dirac particles.
An enhancement in cross-sections 
induced by atomic effects,
as depicted in Figure~\ref{fig::dNdT}a,
has recently been identified~\cite{nuemai,numq}.
The known ratios of peaks at discrete binding 
energies provide
smoking-gun signatures for positive observations.

It follows from Figure~\ref{fig::dNdT}a 
that experimental studies on 
$\munu$ and $q_{\nu}$
should focus on $\rm{T < 10~ \keVee}$.
At benchmark experimental sensitivities
and with comparable exposure as the 
GEMMA experiment~\cite{gemma}, 
the potential reaches are
$\munu \sim 2 \times 10^{-11} ~ \mub$ 
and $\delta _{\rm Q} \sim 6 \times 10^{-14}$, 
where $\mub$ is the Bohr magneton.

In addition, it was recognized~\cite{NRnuCoh} 
that the $\munu$-induced interaction 
with matter would have
a pronounced enhancement in cross-section,
manifesting as measurable peaks
when the initial-state neutrinos
are non-relativistic.
The experimental signatures require
good-resolution and low-threshold
measurements for which Ge detectors 
would be optimal.

\subsection{Neutrino Nucleus Coherent Scattering}

The elastic scattering between a neutrino and a 
nucleus ($\nu$N)~\cite{texonoprogram,nuNcoh} 
\begin{equation}
\rm{
\nu ~ + ~ N ~ \rightarrow ~
\nu ~ + ~ N
}
\end{equation}
is a fundamental SM interaction 
that has never been observed~\cite{nuNcohexpt}.
It probes coherence effects in electroweak interactions~\cite{nuNalpha}, 
and provides a sensitive test for physics beyond SM.
The coherent interaction plays an important role 
in astrophysical processes
and constitutes the irreducible background to
the forthcoming generation of dark matter experiments.
Coherent neutrino scattering may provide new approaches to 
the detection of supernova neutrinos
and offer a promising avenue towards 
a compact and transportable neutrino detector 
capable of real-time monitoring of nuclear reactors.

The maximum nuclear recoil energy for a Ge target (A=72.6) 
due to reactor $\nuebar$ is about 2~${\rm keV_{\rm nr}}$.
The quenching factor (QF, discussed in Section~\ref{sect::qf}), 
is about 0.2 for Ge in the $< {\rm 10~ keV_{\rm nr}}$ region. 
Accordingly, the maximum measurable energy 
for nuclear recoil events in Ge due to reactor
$\nuebar$ is about 300~$\eVee$. 
The typical differential spectrum 
and the integral event rate as a function of detection threshold 
are given in Figures~\ref{fig::dNdT}a\&b, respectively.
At benchmark sensitivities, 
the expected rate is of $\mathcal{O}{\rm ( 10~kg^{-1}day^{-1} )}$ 
with a signal-to-background ratio $>$50.
Improvement of the lower reach of detector sensitivity
is therefore crucial for such experiments.


\begin{figure}
{\bf (a)}\\
\includegraphics[width=8.5cm]{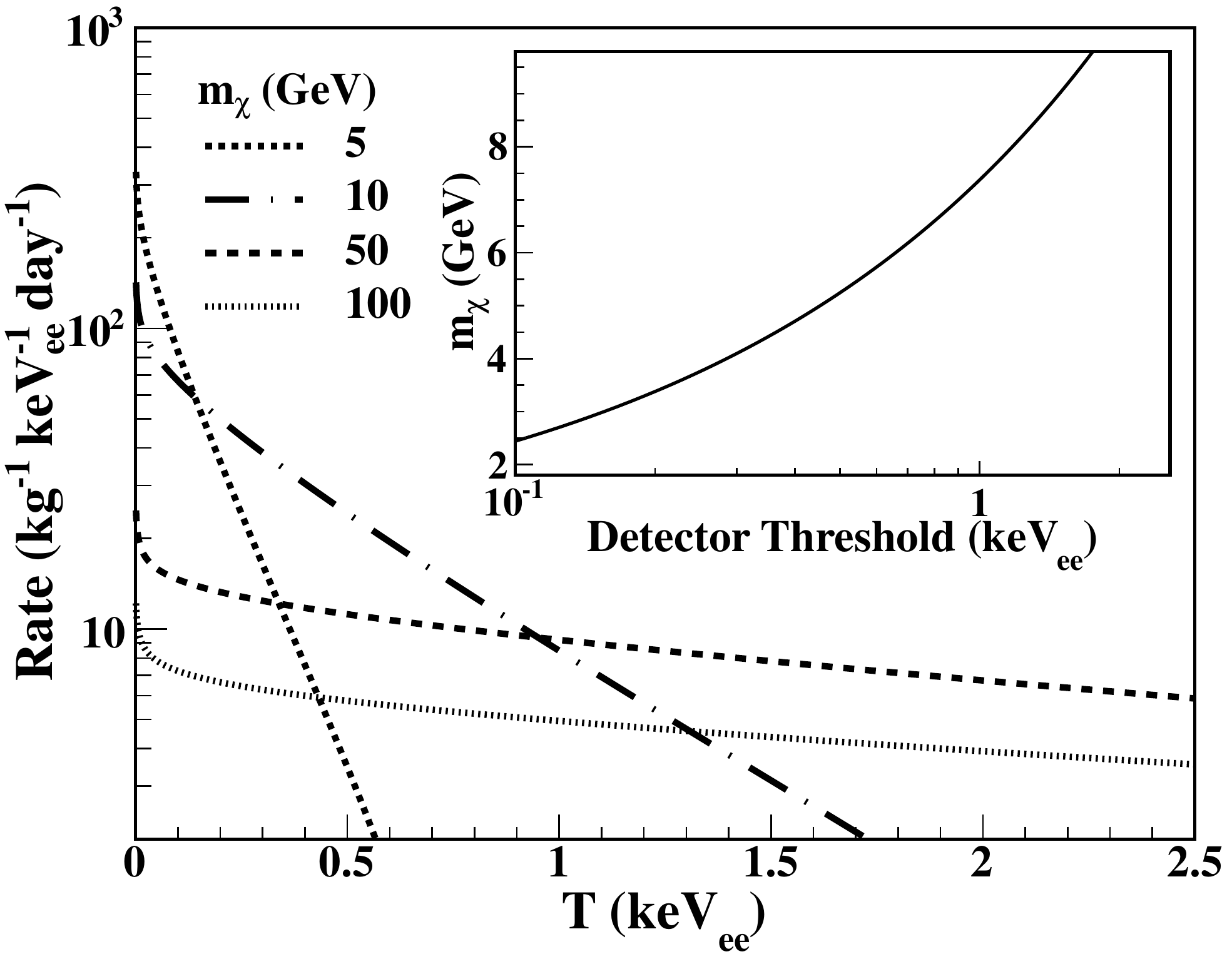}\\
{\bf (b)}\\
\includegraphics[width=8.5cm]{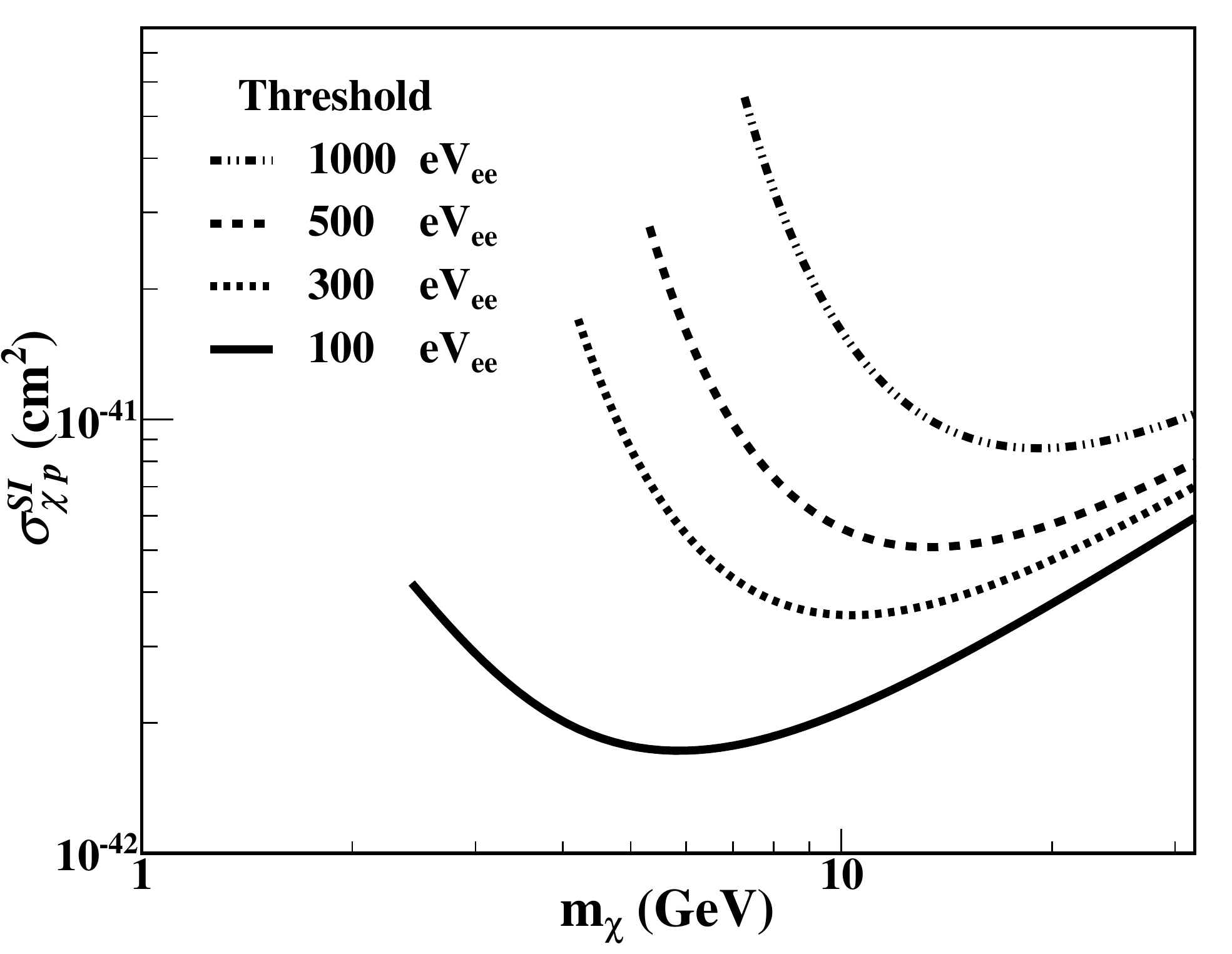}
\caption{
(a)
Recoil spectra for $\chi$-Ge interactions at a cross-section
of $\rm{10^{-40}~ cm^2}$, at various $\rm{m_\chi}$ values.
The lower bounds of $\rm{m_\chi}$
as a function of physics threshold are shown in the inset,
assuming 1~kg-yr of data and a background level of $1~\cpkkd$.
Quenching effects of nuclear recoils are taken into account.
(b)
Sensitivity reach in $\chi$-proton
spin-independent cross-section ($\sigma^{SI}_{\chi p}$)
of the same configuration at different 
detector thresholds, showing
the relative improvement in cross-section 
as a function of $\rm{m_\chi}$.
}
\label{fig::cdmplot}
\end{figure}


\subsection{Dark Matter Searches}

Weakly Interacting Mass Particles (WIMPs, denoted by $\chi$) 
are leading dark matter candidates~\cite{cdmpdg12}.
The elastic scattering between WIMPs 
and the nuclei
\begin{equation}
\rm{
\chi ~ + ~ N ~ \rightarrow ~
\chi ~ + ~ N 
}
\end{equation}
is the favored channel in direct dark matter
search experiments.
Consistency with observations on 
formation of cosmological structures
requires that WIMPs
should be massive and non-relativistic.
In addition, interactions between WIMPs and matter
may be both spin-independent and spin-dependent.
Requirements of experimental studies on WIMPs
are similar to those on neutrino-nucleus coherent scattering,
in which a low detector threshold plays a crucial role.

As illustrations, the nuclear recoil 
spectra for Ge with $\rm{\sigma = 10^{-40} ~ cm^2}$
at various WIMP masses ($\mwimp$)
are displayed in Figure~\ref{fig::cdmplot}a.
A reduction in detector threshold opens
a new observation window for low-mass WIMPs. 
A germanium detector with 100~$\eVee$ threshold would allow
light WIMPs with $\rm{m_\chi}$ down to 2~GeV to be 
probed~\cite{texono2009,texono2013,cogent,cdex0,cdex1,malbek}.
The sensitivity reach on $\rm{m_\chi}$ 
with \mbox{1~kg-yr} of exposure as a
function of threshold at a background
of 1~$\cpkkd$ is illustrated 
in the inset of Figure~\ref{fig::cdmplot}a.
Moreover, a lower threshold allows a wider range of WIMPs to contribute
in an observable interaction and hence results
in better sensitivities for all values of $\mwimp$,
as shown in Figure~\ref{fig::cdmplot}b.



\begin{figure}
{\bf(a)}\\
\includegraphics[width=8.0cm]{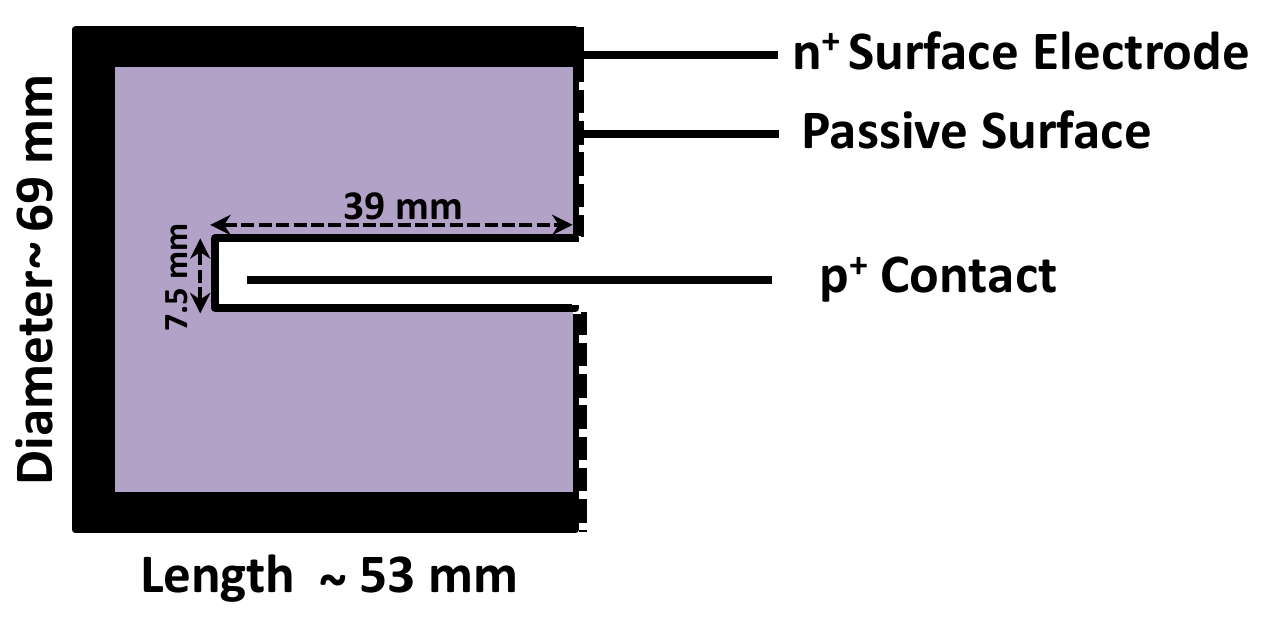}\\
{\bf(b)}\\
\includegraphics[width=8.0cm]{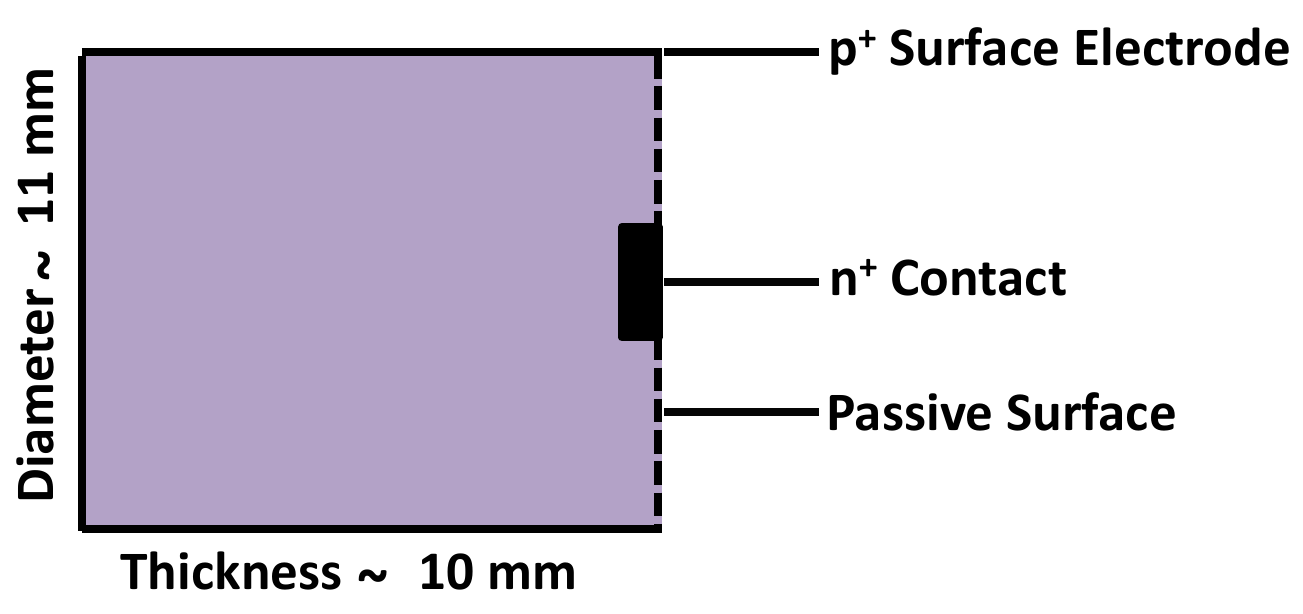}\\
{\bf(c)}\\
\includegraphics[width=8.0cm]{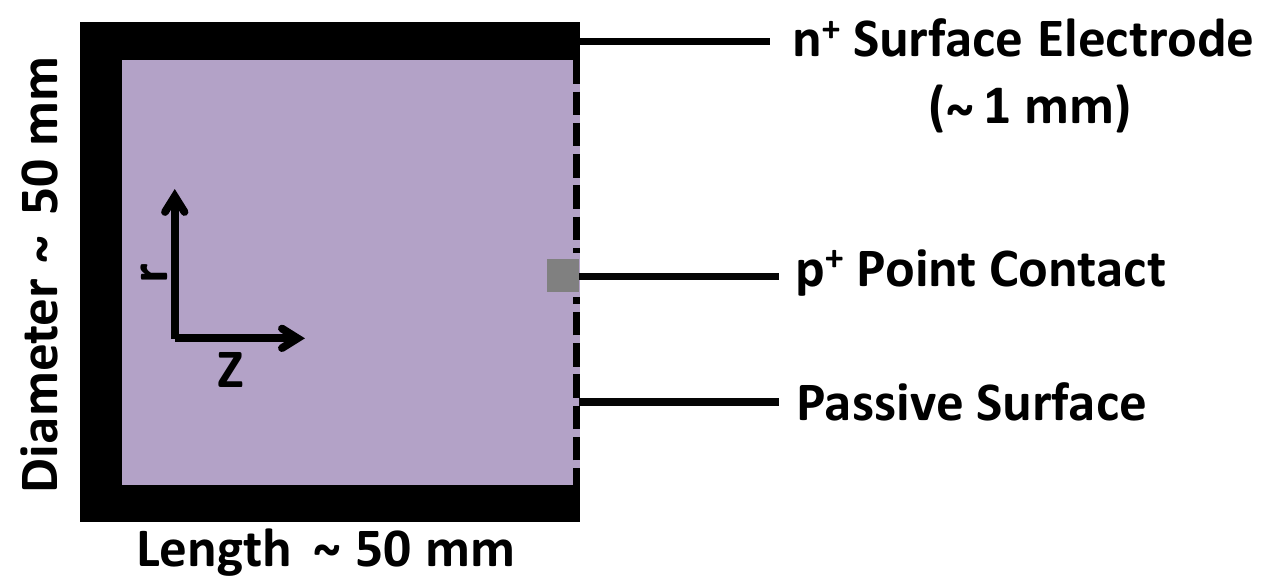}\\
{\bf(d)}\\
\includegraphics[width=8.0cm]{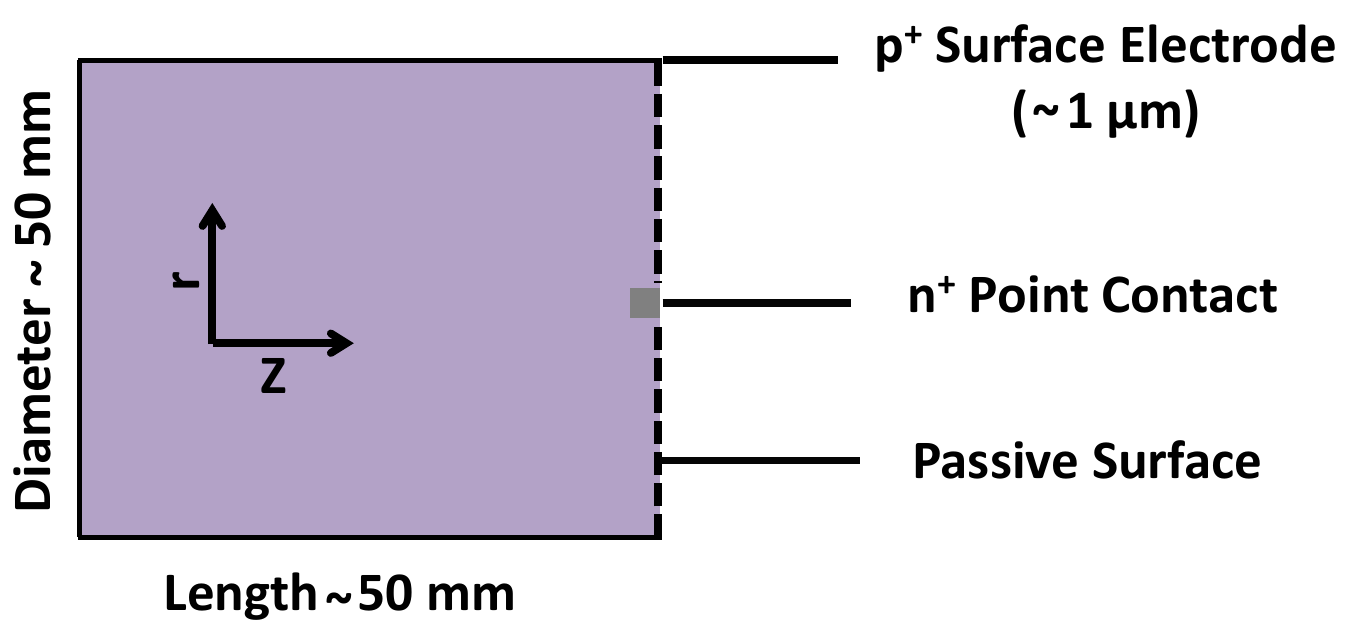}
\caption{
Schematic crystal configuration of
Ge detectors discussed in this article:
(a) CoaxGe with 1~kg mass,
(b) ULEGe with 5~g modular mass,
(c) pPCGe with 500~g mass, and
(d) nPCGe with 500~g mass.
}
\label{fig::subkevge}
\end{figure}



\begin{figure}
{\bf (a)}\\
\includegraphics[width=8.cm]{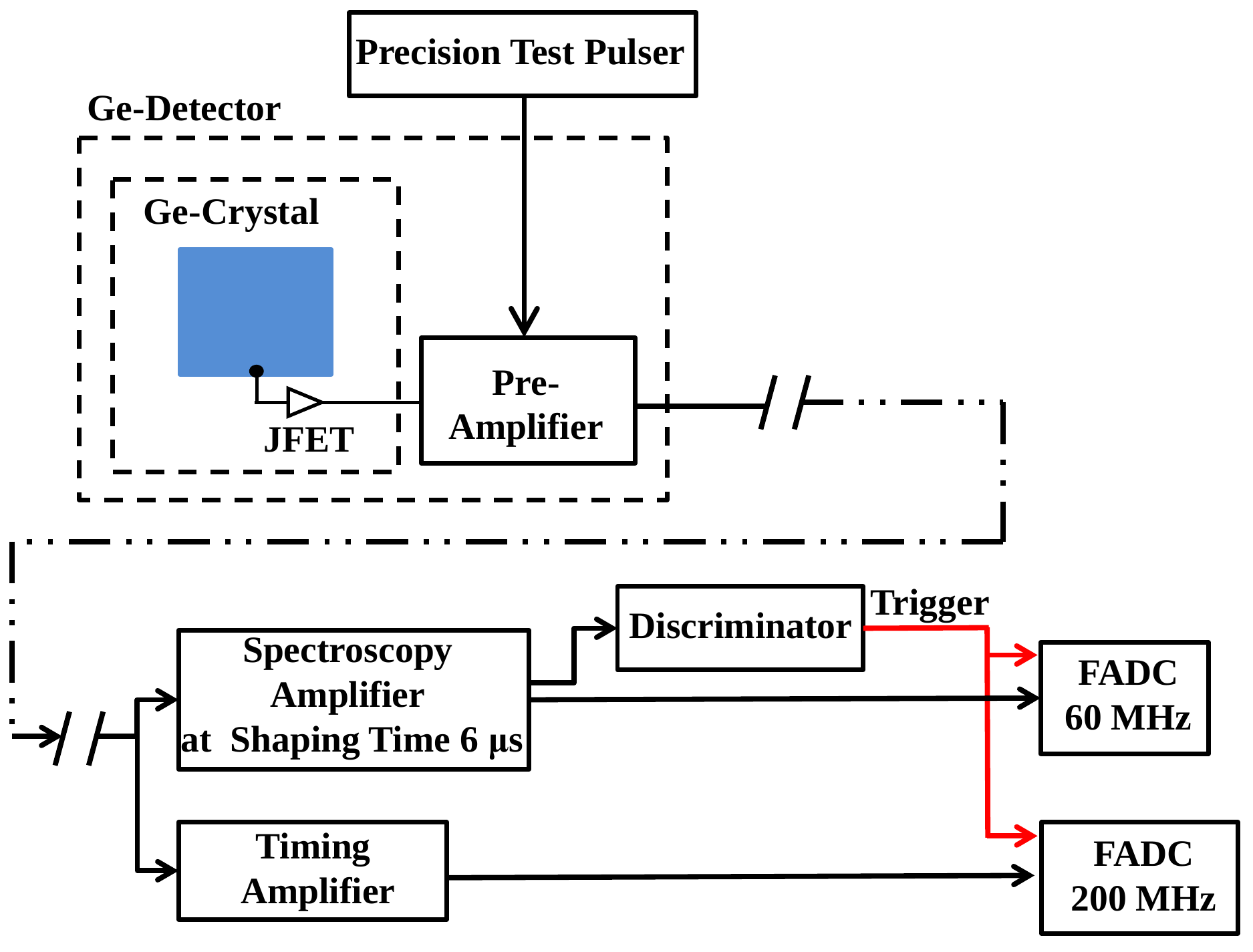}\\
{\bf (b)}\\
\includegraphics[width=8.5cm]{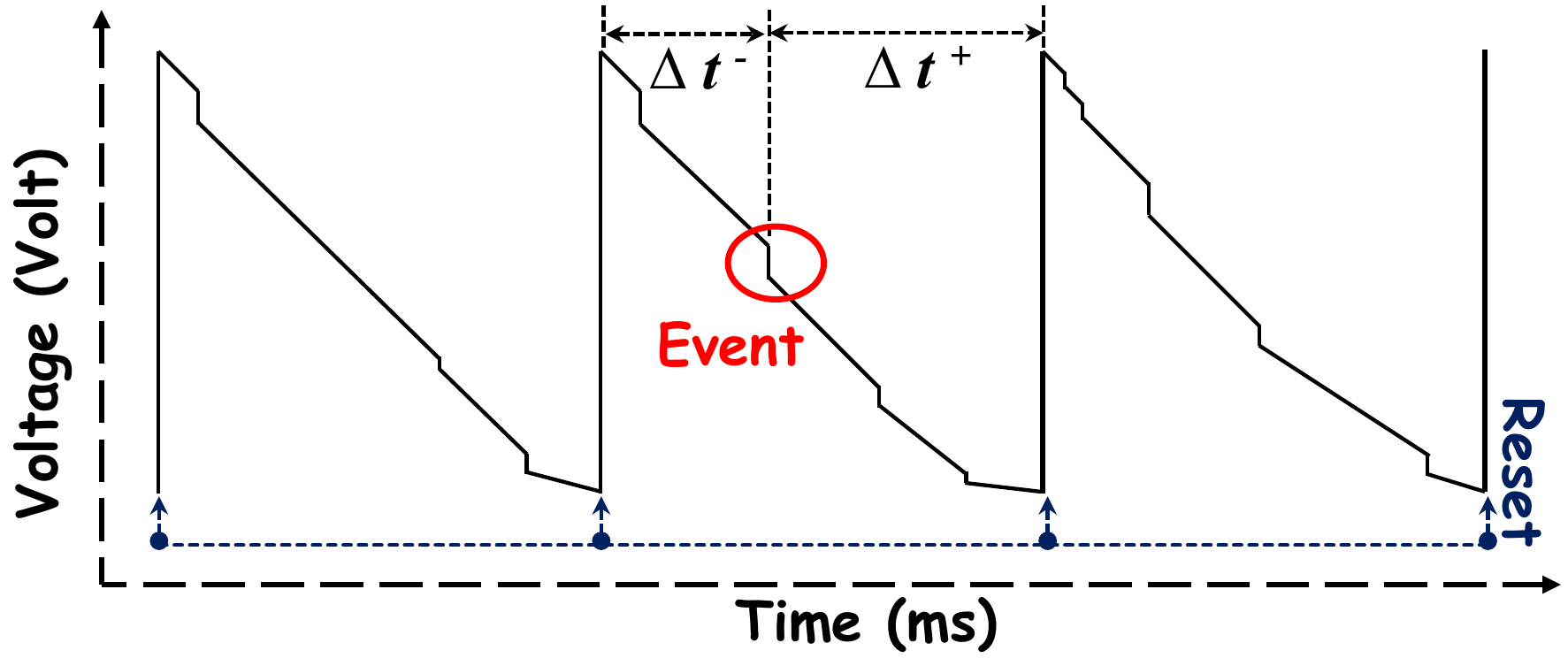}\\
{\bf (c)}\\
\includegraphics[width=8.5cm]{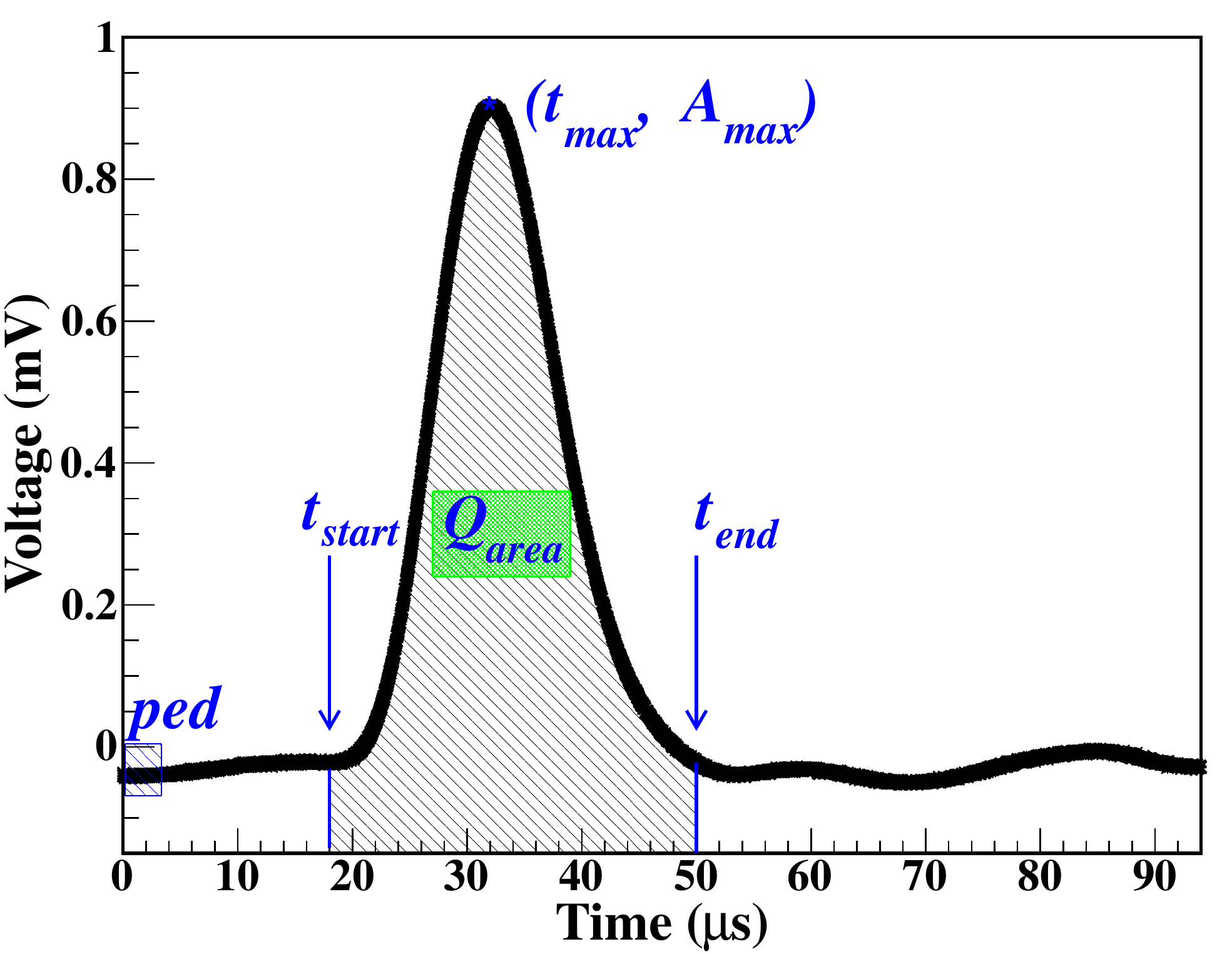}
\caption{
(a)
Schematic diagram of the DAQ system
for detector performance characterization
reported in this work. The system in the experiment
at KSNL includes readout of NaI(Tl) anti-Compton
detector and cosmic-ray veto scintillator panels.
(b)
Schematic drawing of raw preamplifier signals
as recorded with an oscilloscope. The RESET
amplitude and time interval of
different detectors are summarized
in Table~\ref{tab::summary-performance}.
(c)
Typical $\sasix$ pulse at 6~$\mu$s shaping time.
Various key parameters for analysis
and calibration purposes are shown 
in both (b) and (c).
}
\label{fig::pulseshape}
\end{figure}



\section{Sub-keV Germanium Detector}

\subsection{Detector and Readout}

Data taken with four Ge detectors are adopted in this study.
The detectors and their characteristics
are listed as follows, 
while their respective sensor electrode schematics are depicted 
in Figures~\ref{fig::subkevge}a,b,c\&d.
\begin{enumerate}
\item Conventional 
coaxial p-type high purity Ge detector (CoaxGe)
with 1~kg in mass,
used as target detector in Ref.~\cite{texonomunu};
\item 4-element array of n-type Ge detector
with 5~g modular mass, a pilot 
``ultra-low-energy'' germanium detector (ULEGe)
used in Refs.~\cite{texono2009,cdex0};
\item p-type point-contact Ge detector (pPCGe) 
with 500~g mass, similar in functionalities but not in detector mass
to those used as target detectors in 
Refs.~\cite{texono2013,cdex1}; and
\item n-type point-contact Ge detector (nPCGe)
with 500~g mass, used as calibration detector in 
Refs.~\cite{texono2013,bsel2014}.
\end{enumerate}

All the detectors have been procured commercially.
CoaxGe and ULEGe are conventional ``catalog-item''
detectors\footnote{Canberra Meriden, USA.}
serving for control and comparison purposes.
Both pPCGe and nPCGe are custom-designed 
and fabricated\footnote{Canberra Lingolsheim, France.},
and are the primary focuses of this work.

The concept of point-contact Ge detectors
was proposed and the first nPCGe with 800~g mass
was constructed in the 1980s~\cite{lukepcge}, followed
by recent realization of pPCGe~\cite{cogentppcge}
for low-threshold, low-background experiments.
Sub-keV threshold is realized using ULEGe, pPCGe and nPCGe 
through optimizing the detector configurations 
that reduces the output capacitance to $\sim$pF, 
and through improving the front-end ``JFET'' electronics.
The subsequent discussions 
apply to these three detectors,
except those on differentiation of surface and bulk
events in Section~\ref{sect::bsppcge}i, which
are relevant only to pPCGe.
The performance parameters of the four detectors 
in this study are summarized in Table~\ref{tab::summary-performance},
while those of earlier benchmark projects 
with Ge detectors in which threshold is crucial
are listed in Table~\ref{tab::survey} for comparison.

A schematic diagram of the readout scheme for
characterizing Ge detectors 
is illustrated in Figure~\ref{fig::pulseshape}a.
Signals from Ge-crystal sensors 
are first amplified by 
front-end JFETs\footnote{Custom-built 
for low electronics noise, Canberra Lingolsheim.
Please contact the company for technical information.}
located in the vicinity of the Ge diodes.
Outputs are fed to 
reset preamplifiers\footnote{Model PSC954, Canberra Lingolsheim.} 
placed $\sim$30~cm away. 
The typical output as observed on an oscilloscope
is displayed in Figure~\ref{fig::pulseshape}b.
The saw-tooth waveform exemplifies
the timing structures of ``RESETs'' 
issued after a fixed time interval or 
when charge deposition in the detector exceeds
a pre-set value (for example, with a direct cosmic-ray event).
The steps in between RESETs
represent physics signals whose
energy is proportional to the step size.


\input{table_summary_performance.tex}


\input{table_survey}


As depicted in Figure~\ref{fig::pulseshape}a,
the preamplifier signals are 
further processed by both shaping and timing
amplifiers\footnote{Canberra 2026 and 2111, respectively.}. 
Output from the timing amplifier (TA) preserves 
rise-time information 
for distinguishing bulk and surface 
events (Section~\ref{sect::bsppcge}).
The shaping amplifier signals at 6~$\mu$s 
shaping time ($\sasix$) are optimized for energy measurement. 
The discriminator output provides the trigger instant 
for data acquisition (DAQ).
Pulses from TA and $\sasix$ are digitized by 200 and 60~MHz 
flash analog-to-digital 
converters\footnote{National Instruments 
PXI 5105 and PXI 5124, respectively.},
respectively. 
A typical $\sasix$ pulse at 4.2~$\keVee$ for pPCGe
is shown in Figure~\ref{fig::pulseshape}c.

\subsection{Data Samples}

\begin{figure}
{\bf (a)}\\
\includegraphics[width=8.5cm]{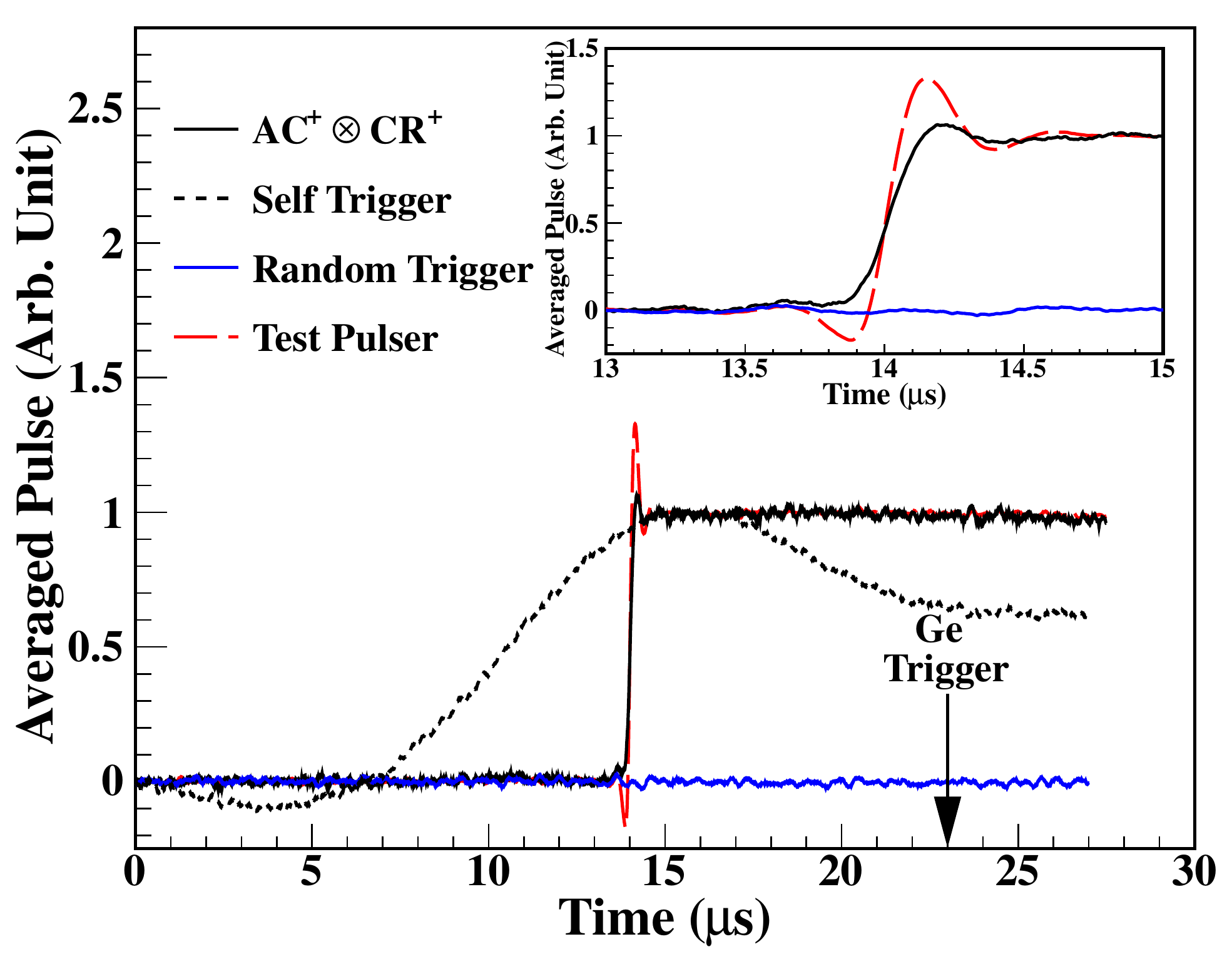}\\
{\bf (b)}\\
\includegraphics[width=8.5cm]{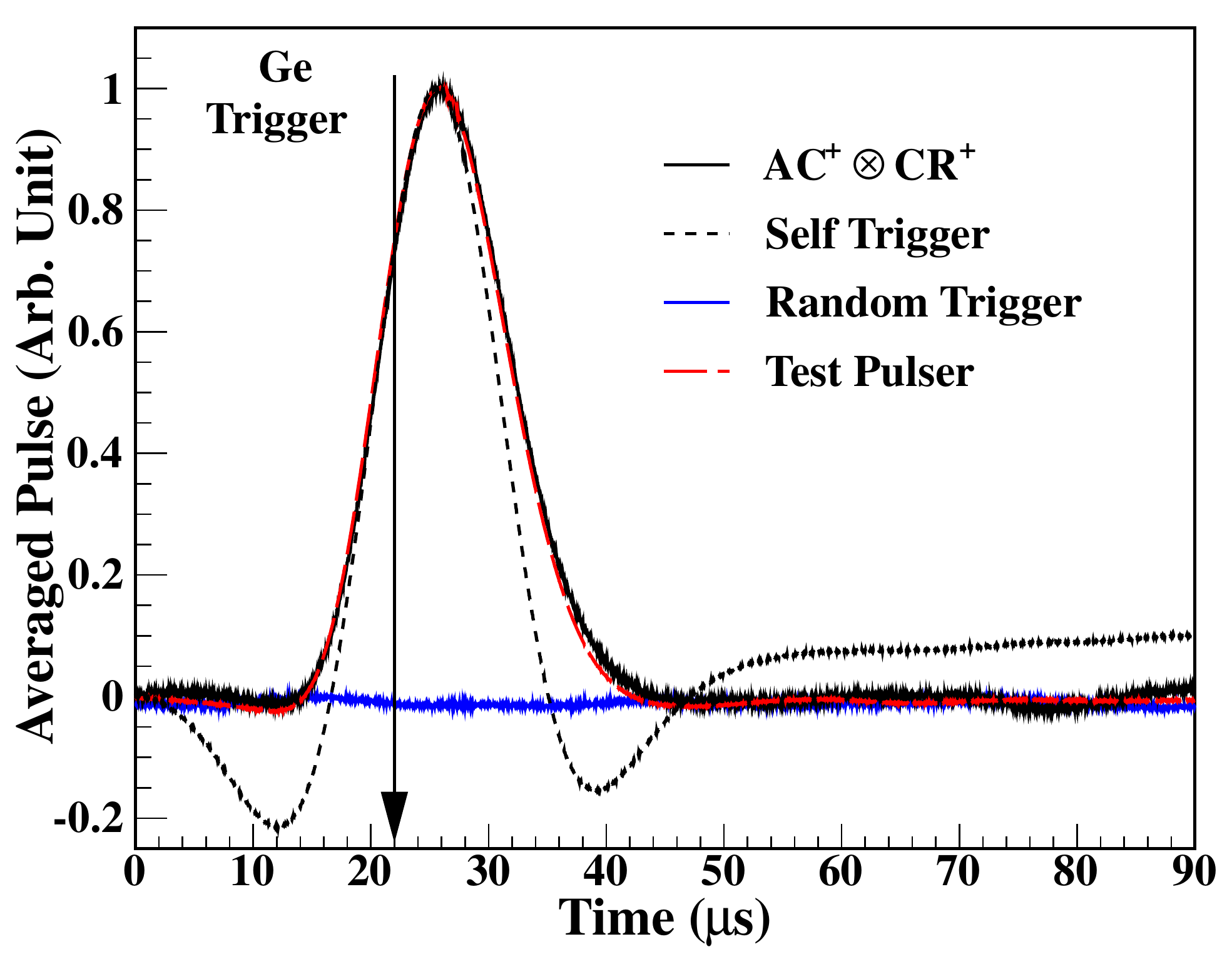}
\caption{
Comparison of averaged pulse shapes 
from (a) timing amplifier TA and  
(b) shaping amplifier $\sasix$ 
for events due to 
random-trigger (blue solid),
self-trigger pedestal electronic noise (black dashed),
test-pulser (red dashed) and physics interactions (black solid).
Data from pPCGe are used as illustrations.
The selected events except random-trigger ones 
are of effective energy near noise-edge 
($\sim 300~\eVee$ in this example).
Their amplitude is normalized to unity in the display,
except for random-trigger events whose normalization 
follows that of self-trigger events.
The physics samples are from 
bulk events tagged with ``AC$^+$$\otimes$CR$^+$'' 
and after basic filters of Section~\ref{sect::basicfilters}.
The trigger instants defined by Ge-$\sasix$ signals are shown.
The physics and test-pulser events 
are identical in $\sasix$ but differ in TA.
The physics and self-trigger noise events 
show different profiles in both amplifiers
and are therefore, {\it in principle},
distinguishable.
}
\label{fig::averagedshape}
\end{figure}

Characterizations of detector performance 
are carried out at our home-based laboratories.
Physics measurements were made 
with the detectors placed inside 
the low-background facilities~\cite{texonomunu,texononuecsi}
at Kuo-Sheng Reactor Neutrino Laboratory (KSNL)
which has about 30~meter-water-equivalent overburden.
Data of the Ge detector were recorded 
in conjunction with 
an NaI(Tl) anti-Compton (AC) detector and 
a cosmic-ray (CR) veto scintillator 
array~\cite{texonomunu,texono2009,texono2013}.
The NaI(Tl)-AC detector has a mass of 38.3~kg
and a well-shaped geometry to enclose
the Ge detectors.
The entire setup is housed in a shielding structure. 
A background level of $\rm{ 1 - 10 ~ \cpkkd}$
at the $\keVee$-range was achieved.
The averaged pulse shapes of TA and $\sasix$ output
due to different event categories are displayed
in Figures~\ref{fig::averagedshape}a\&b, respectively.
The trigger instant from discriminator output of 
$\sasix$ is superimposed. 
It is defined by the time
when the amplitude of $\sasix$ signals ($\eamp$)
surpasses certain pre-set discriminator level ($\Delta$). 
The trigger is therefore issued 
with a time delay of about 10~$\mu$s at low energy
relative to the prompt signals from TA, AC and CR.

Events at KSNL can be categorized by
``AC$^{-(+)}$$\otimes$CR$^{-(+)}$'', where
the superscript $-(+)$
denotes anti-coincidence(coincidence)
with Ge signals.
Physics events are those with genuine energy depositions
in the Ge detectors.
 
A pure sample of physics events for calibration purposes 
can be collected by  ``AC$^+$$\otimes$CR$^+$'' tag
where {\it all} three detectors are in coincidence.
Candidate events of neutrino or WIMP-induced interactions
are uncorrelated with other detectors and are therefore
extracted from ``AC$^-$$\otimes$CR$^-$'' tags.


At signal amplitude comparable to 
pedestal fluctuations,
the AC$^-$$\otimes$CR$^-$ triggers are
mostly due to self-trigger electronic noise.
Positive fluctuations of these
noise events in the shaped pulses 
of Figure~\ref{fig::averagedshape}b
at the trigger instant are accompanied with
negative fluctuations prior to and ${\rm \sim 20~\mu s}$
after the triggers. 
The corresponding TA pulse is shown 
in Figure~\ref{fig::averagedshape}a.
It has a slow rise-time and 
also decreases in amplitude when
the maximum is reached.
This characteristic feature is expected because the
self-trigger noise events originate from pedestal 
fluctuations rather than genuine energy depositions
due to physics interactions at the detectors.

Self-trigger pedestal noise events 
are therefore, {\it in principle},
distinguishable from physics events by their
different pulse shapes.
However, it is technically challenging to devise
efficient pulse-shape discrimination techniques 
at the event-by-event basis for physics signals 
with amplitude comparable to that of pedestal fluctuations.
Developing advanced analysis algorithms 
for this task is beyond the scope of this report
but continues to be our research efforts. 

Random-trigger events are taken from
random sampling of the pedestal baseline
for monitoring and calibration purposes. 
They are useful for 
quantifying pedestal fluctuations,
defining zero-energy offset
and measuring DAQ dead-time and various 
efficiency factors.

Test-pulser events are produced by 
a precision pulse 
generator\footnote{Precision Test Pulser National Instruments PXI 5412.}
fed to the preamplifier. 
Pulser signals probe the response of 
electronics system independent of the 
electron drifting effects for physics signals.
As demonstrated in Figure~\ref{fig::averagedshape}b,
pulser and physics events have identical profiles
in $\sasix$. 
Accordingly, pulser events can be used for
studying the detector energy response (Section~\ref{sect::Qresponse})
and measuring trigger efficiency (Section~\ref{sect::daqeff}),
because both measurements are defined by the $\sasix$ pulses.
However, as depicted in Figure~\ref{fig::averagedshape}a,
pulser signals have faster rise-time than physics events
(bulk samples tagged with ``AC$^+$$\otimes$CR$^+$'') 
in their TA output. 
It would be unjustified to adopt test-pulser events 
for efficiency calibrations that involve 
pulse-shape analysis of TA signals, 
such as in the identification of 
bulk/surface events (Section~\ref{sect::bsppcge}i).

We note that different test-pulser models,
or the same model under different settings, 
give different TA output. 
The pulser profile of Figure~\ref{fig::averagedshape}a
is due to a setting that gives the 
closest match to that of physics events.
Detailed understanding and optimization of
pulser parameters for making TA output 
compatible with physics signals
are beyond the scope of this work,
but represent a relevant direction of future research. 

\begin{figure}
\includegraphics[width=8.5cm]{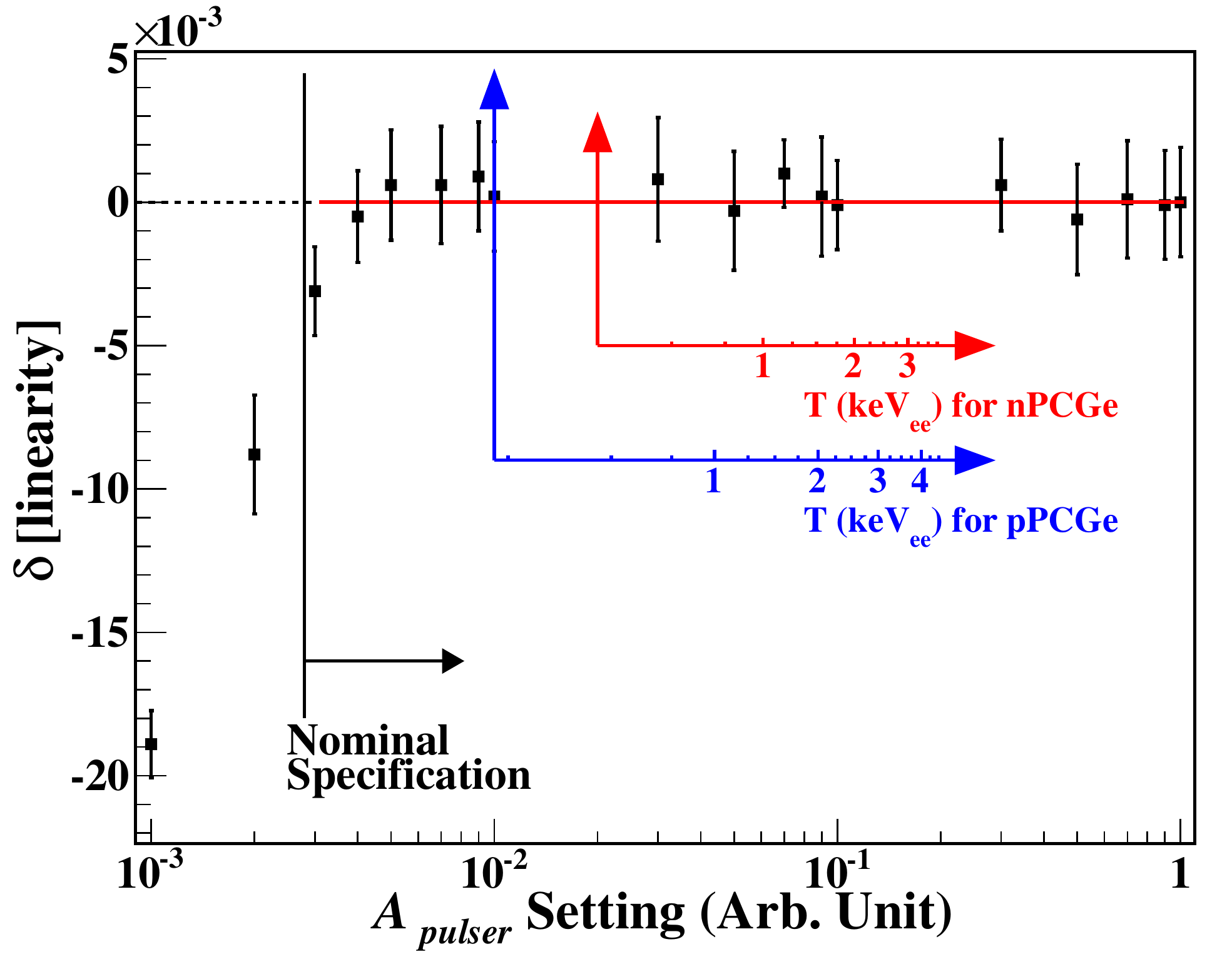}\\
\caption{
Direct measurement of linearity of 
test-pulser output relative
to input setting ($A_{Pulser}$).
Superimposed are the nominal range specified by
the manufacturer.
Corresponding energy scales in $\keVee$ above 
electronic noise-edges of
pPCGe and nPCGe are also shown.
Deviations from linearity are described by 
${\rm \delta [linearity]}$, which
is the fractional deviations of 
measured amplitude from the nominal 
values. 
}
\label{fig::pulser}
\end{figure}

The linearity of test-pulser 
output with respect to input setting ($A_{Pulser}$)
is verified by direct measurement of raw pulser signals,
as depicted in Figure~\ref{fig::pulser}. 
Superimposed are the energy-equivalent settings for
electronic noise-edges of pPCGe and nPCGe, 
as well as the nominal range specified
in factory data-sheet, demonstrating the validity
of pulser measurements for our current studies.


\begin{figure}
{\bf (a)}\\
\includegraphics[width=8.5cm]{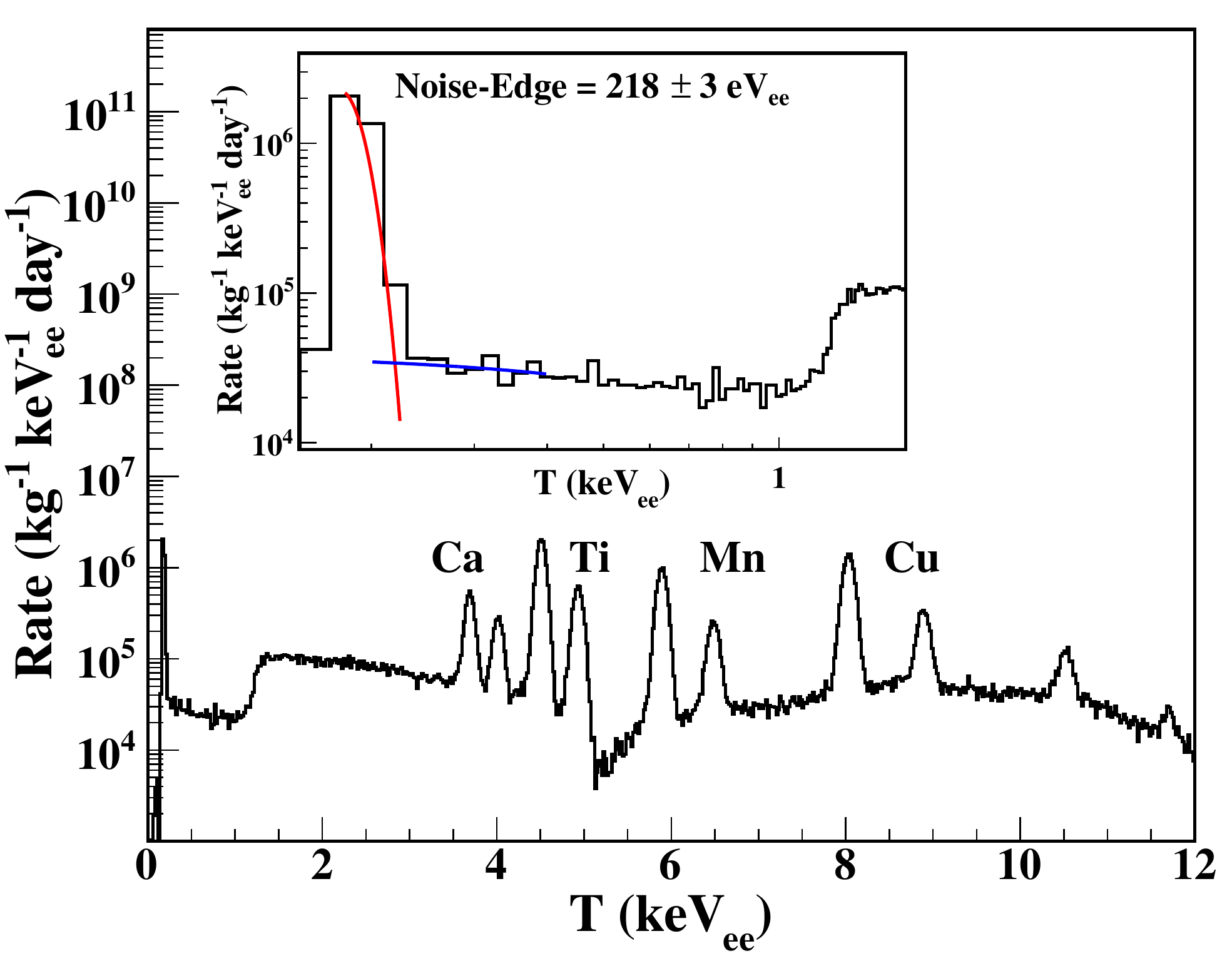}\\
{\bf (b)}\\
\includegraphics[width=8.5cm]{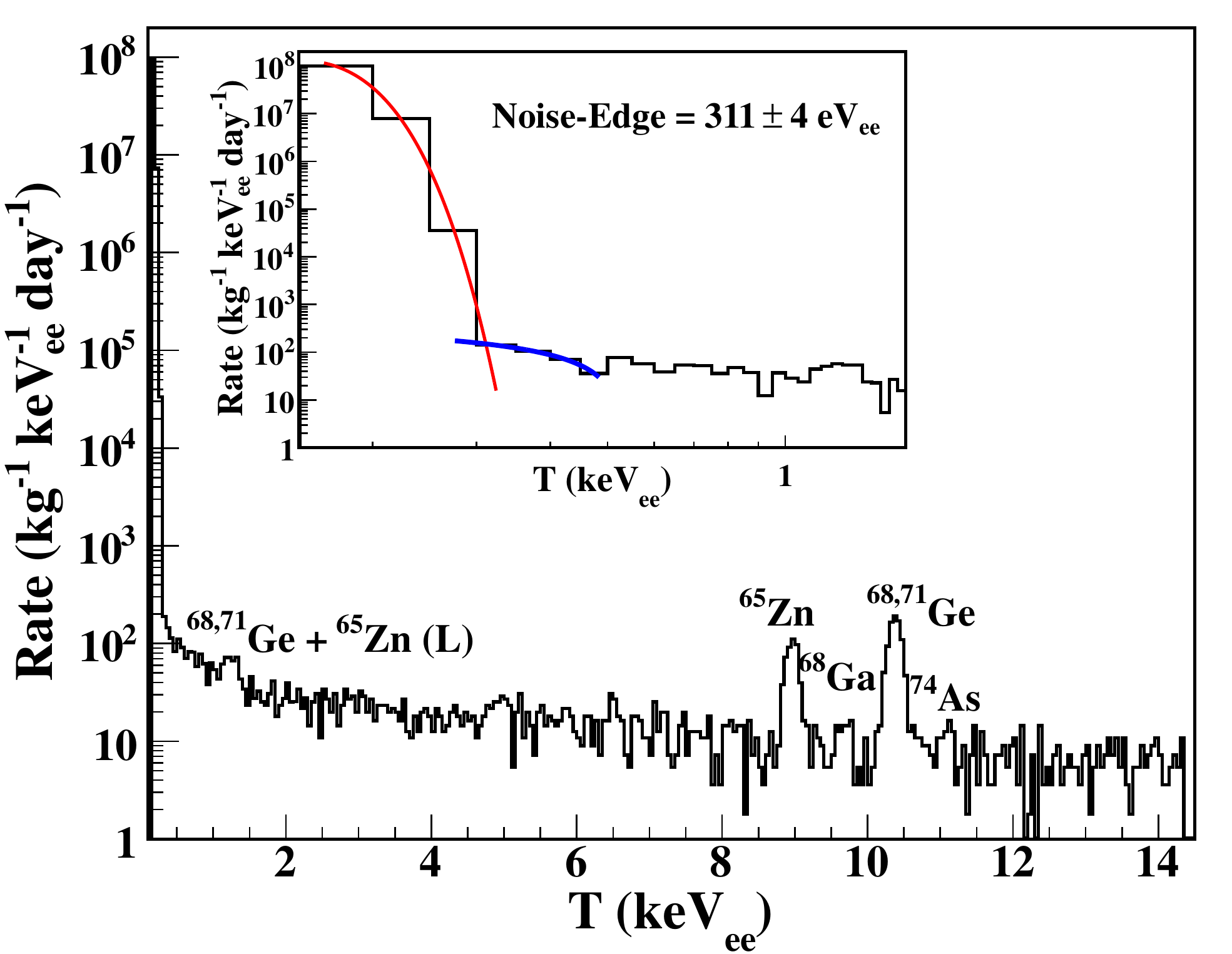}\\
{\bf (c)}\\
\includegraphics[width=8.5cm]{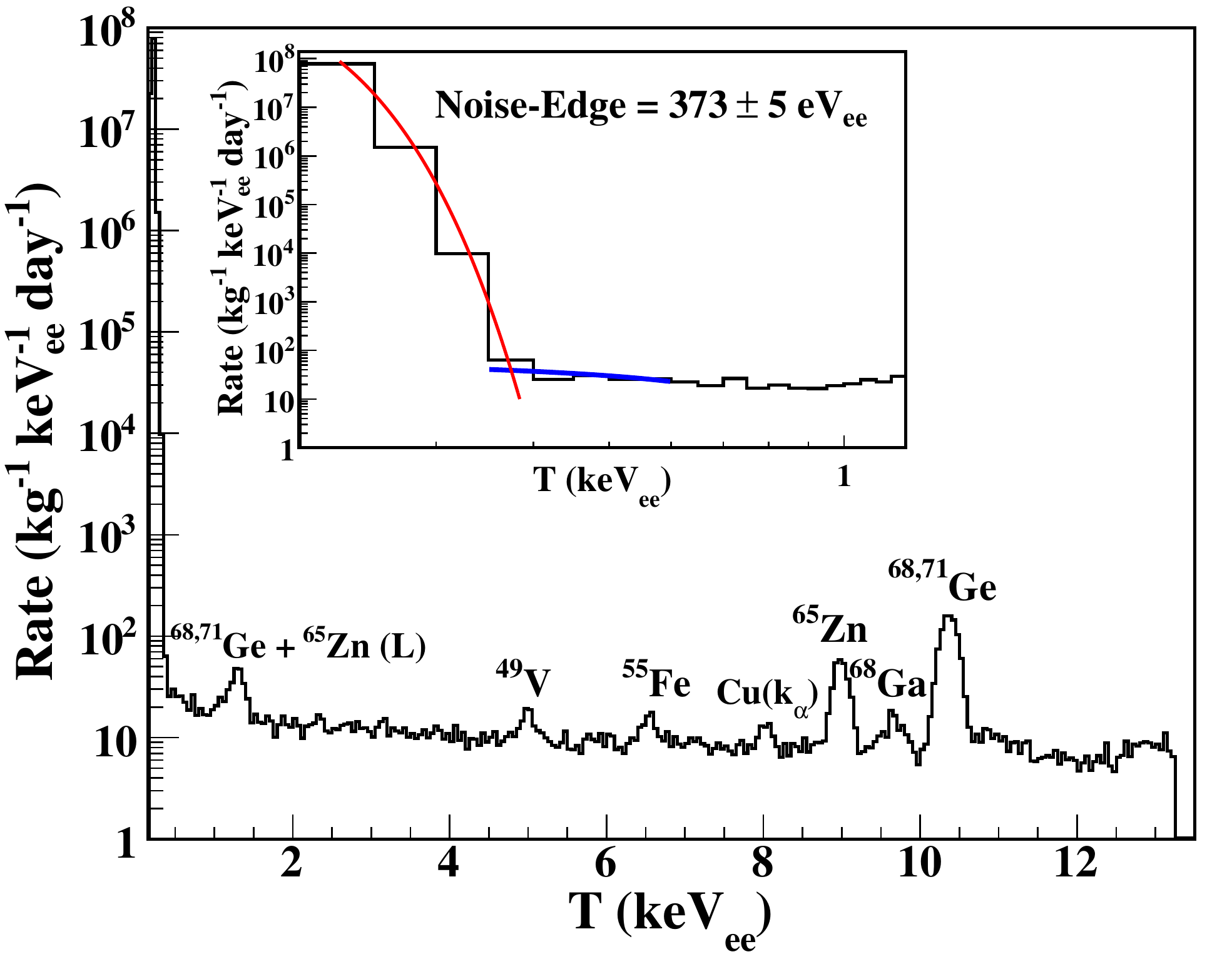}
\caption{
Typical
(a) ULEGe, (b) pPCGe and (c) nPCGe spectra 
showing X-ray peaks and noise-edges.
The lines in all cases are used in energy calibration.
ULEGe is an n-type detector with a thin surface layer. 
It is also equipped with a thin cryostat
window which allows detection of external X-ray lines in (a).
The two PCGe detectors were built for applications with
low count rates and therefore do not have thin windows.
The peaks are due to electron capture in 
isotopes which are cosmogenically activated and
emit X-rays inside the detectors.
Both (b) and (c) are spectra for events after 
basic filters and with AC$^-$$\otimes$CR$^-$ tags.
The noise-edges are illustrated in the insets, 
defined as the energy at which physics signals
would dominate self-trigger
electronic noise events.
}
\label{fig::spectra}
\end{figure}


\subsection{Energy Measurement}
\label{sect::Qresponse}

\subsubsection{Pedestal Noise Characterization}

The pedestal electronic noise of detectors 
are described by ``Pedestal-Noise-Profile-RMS'' ($\amprms$) and 
``Noise-Area-RMS''($\qrms$) 
derived from random-trigger events.  
The RMS (Root-Mean-Square) of pedestal areas integrated 
over $[ t_{start} , t_{end} ]$ is $\qrms$, 
while the bin-by-bin pedestal signal profile distributions
of random-trigger events give $\amprms$. 

There are various merits in choosing
$\amprms$ as a key variable.
As we shall see in subsequent sections, 
various detector behaviors can be described by
universal functions when energy scales 
are expressed in units of $\amprms$.
In addition,
measurements of $\amprms$ in calibrated energy unit
provide a quick-and-valid comparison of
noise levels among different detectors.

The pedestal noise levels of detectors studied 
in this work are summarized 
in Table~\ref{tab::summary-performance}. 

\subsubsection{Energy Estimator}

As illustrated in  Figure~\ref{fig::pulseshape}c, 
$\eamp$ and $\echarge$ of $\sasix$ are 
defined in our studies
as maximal amplitude and 
integrated area within the time window 
$[ t_{start} , t_{end} ]$,
respectively.
The averaged pedestal level is an offset and
is subtracted.
Extensive studies have confirmed that
the selected amplifier shaping time (6~$\mu$s) 
and integration interval (30~$\mu$s) would
produce the best detector performance. 

Conventional applications of Ge detectors 
are at an energy range where $\echarge$ and $\eamp$
are substantially above $\qrms$ and $\amprms$, respectively.
Both parameters have been adopted to estimate
the energy of an event.
When signal amplitude becomes comparable to 
magnitude of electronic noise in low-energy applications,
further investigations are necessary.

For instance, when physics signals far exceed
pedestal noise (as in, for instance,
imaging applications or double beta decay experiments), 
there are established software 
techniques for shaping the fast TA pulse~\cite{softwareshaping}, and
extracting the pulse amplitude information
from the shaped output.
This approach, however, is limited in accuracy,
efficiency and robustness when the signals
are only several times larger than the pedestal noise 
$-$ the range of interest in this work.

Measurements show $\amprms \lesssim \qrms$
for the detectors studied in this work,
as listed in Table~\ref{tab::summary-performance}. 
Pulse amplitude provides a better estimator 
of energy than its area.
Accordingly, the $\sasix$ amplitude $\eamp$ 
is adopted for energy measurement. 
The details of energy response 
are discussed in Section~\ref{sect::energyresponse}.

In addition, the DAQ trigger is issued
when $\eamp$ exceeds 
discriminator threshold $\Delta$. 
Various aspects of the trigger are 
discussed in Section~\ref{sect::trig}.


\begin{figure}
{\bf (a)}\\
\includegraphics[width=8.2cm]{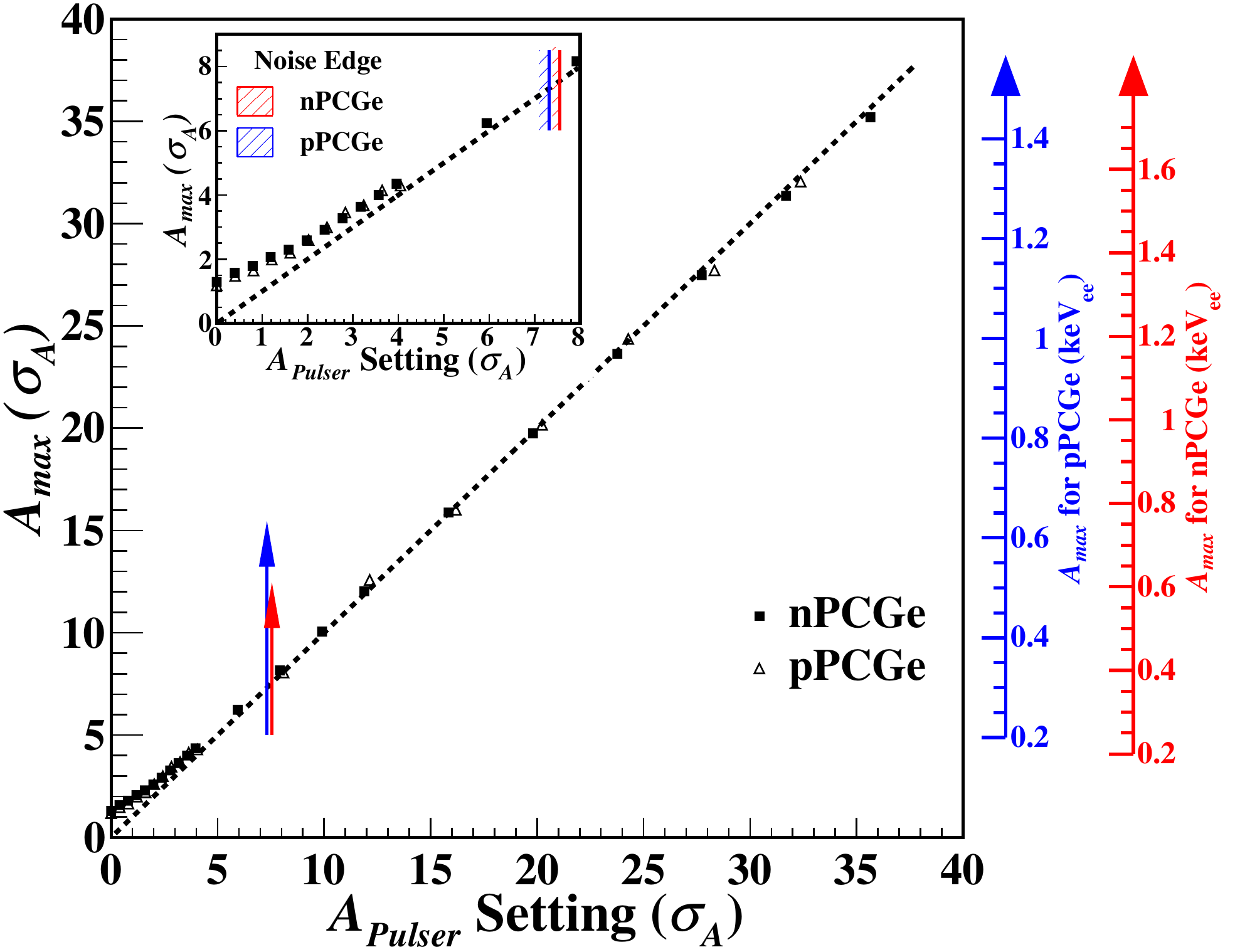}\\
{\bf (b)}\\
\includegraphics[width=8.2cm]{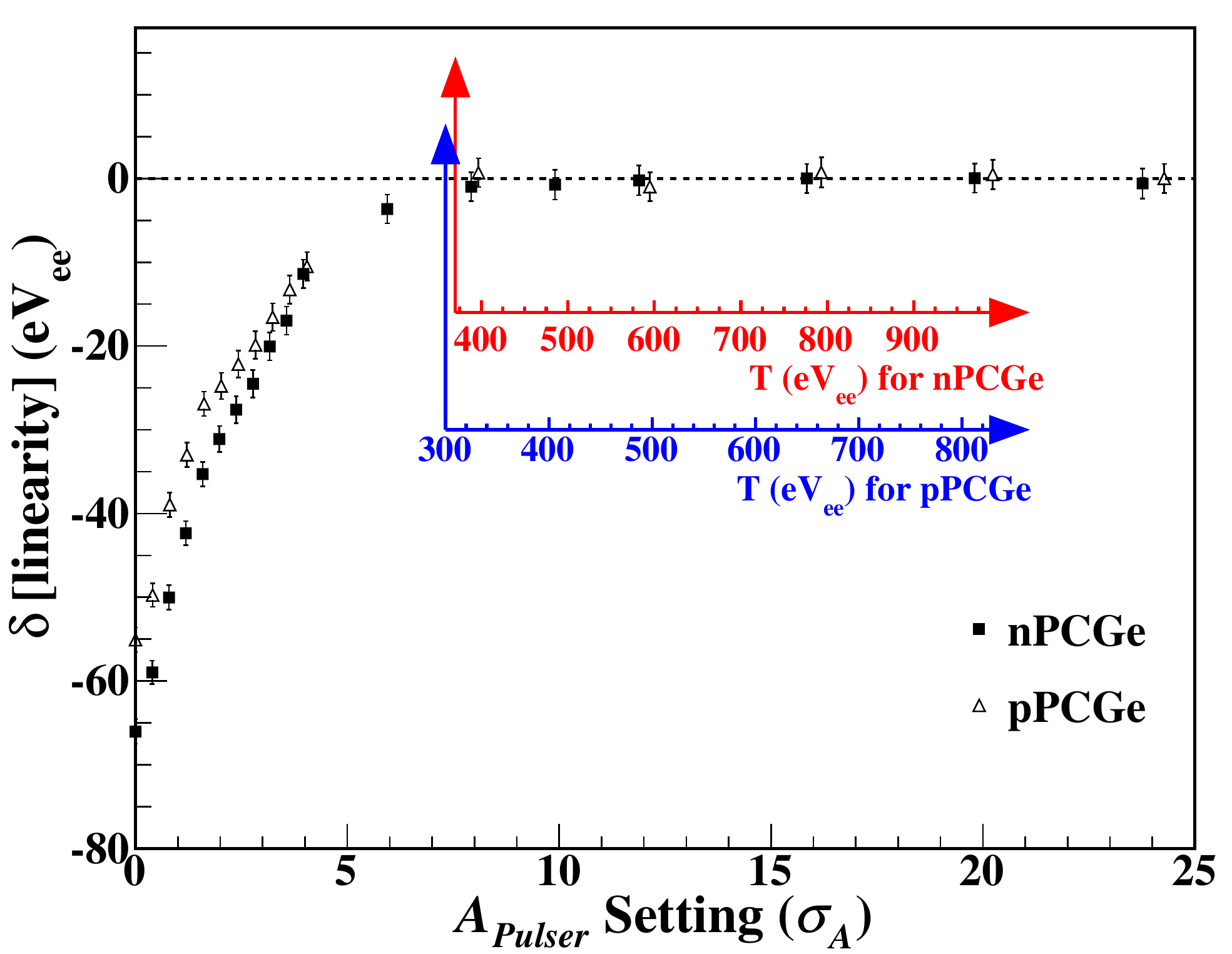}\\
{\bf (c)}\\
\includegraphics[width=8.2cm]{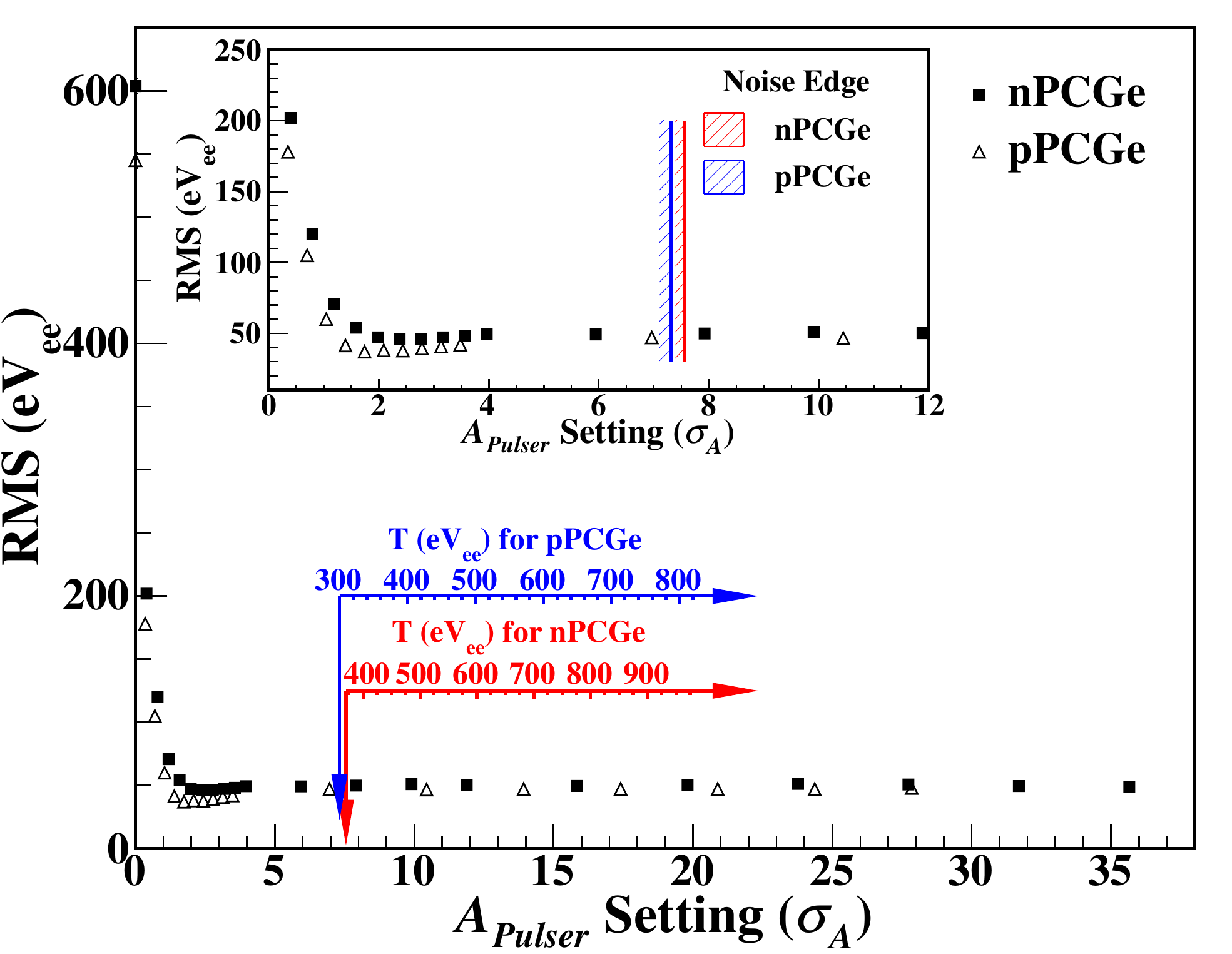}
\caption{
Response of pPCGe and nPCGe
versus energy when the test-pulser amplitude 
is comparable to pedestal
noise fluctuations $-$ 
(a) energy estimator $\eamp$,
(b) deviations from linearity, and
(c) RMS resolution. 
The energy scales 
are in $\amprms$ unit to illustrate universal
behavior, and in $\eVee$ unit 
for specific detectors above their noise-edges.
It can be seen that the 
detector responses are well-behaved
in the physics regions of interest, 
but become anomalous as 
the input approaches zero.
}
\label{fig::response}
\end{figure}


\subsubsection{Energy Calibration}

Shown in Figures~\ref{fig::spectra}a,b\&c
are the measured spectra of
ULEGe, pPCGe and nPCGe, respectively.

The outer boron-implanted dead
surfaces of n-type Ge detectors 
(such as ULEGe, nPCGe) 
are thin ($\mathcal{O}(1~\mu m)$). 
In addition, ULEGe is
equipped with a thin window made of
carbon composite materials. 
It was calibrated by an external X-ray 
generator\footnote{X-Ray Generator: Ampetek Cool-X}
which induces emissions from 
different elements placed close to 
the detector, and gives rise to 
the characteristic X-ray peaks~\cite{texono2009,cdex0}
depicted in Figure~\ref{fig::spectra}a.
The step at $\sim$1~$\keVee$ is due to
attenuation of low-energy photons by the 
thin window.

The {\it in situ} background spectra at KSNL  
for pPCGe and nPCGe are displayed in
Figures~\ref{fig::spectra}b\&c, respectively.
The $\rm{n^+}$ surface electrodes of p-type Ge detectors
are fabricated by lithium diffusion and have
a typical thickness of $\sim$1~mm~\cite{ppcgethickness}.
Therefore, external $\gamma$'s 
with energy less than 50~keV 
are totally suppressed and cannot be used.
Instead, calibration is
provided~\cite{cogent,texono2013,cdex1}
by internal cosmogenic radioactivity,
the strongest of which are peaks from
electron capture of $^{68,71}$Ge
producing Ga X-rays (10.37~keV and 1.29~keV).
This calibration scheme also applies to nPCGe
which does not have a thin entrance window and is
housed in a copper cryostat of 1~mm thickness.

In all the detectors studied, 
no physics-related structures are observed below
$\sim 1 ~ \keVee$, where 
energy calibration is performed using the test-pulser. 
Data were taken with decreasing pulser amplitude.
Pulser measurements at zero-amplitude 
are equivalent to random-trigger events.
The energy scale of the pulser is defined by matching to
the $\gamma$-peaks at high energy. 
As demonstrated in Figure~\ref{fig::pulser},
the intrinsic pulser response is linear at  
the range corresponding to physics events.
The resolution function at sub-keV energy is derived
via extrapolating measured widths of X-ray peaks
at high energy.

The pulser measurements for pPCGe and nPCGe
are displayed in Figure~\ref{fig::response}a.
Polynomial functions provide calibration 
of $\eamp$ into energy unit over the entire range. 
In particular, the response is linear 
at the physics region of interest
above the electronic noise-edge.


\begin{figure}
\includegraphics[width=8cm]{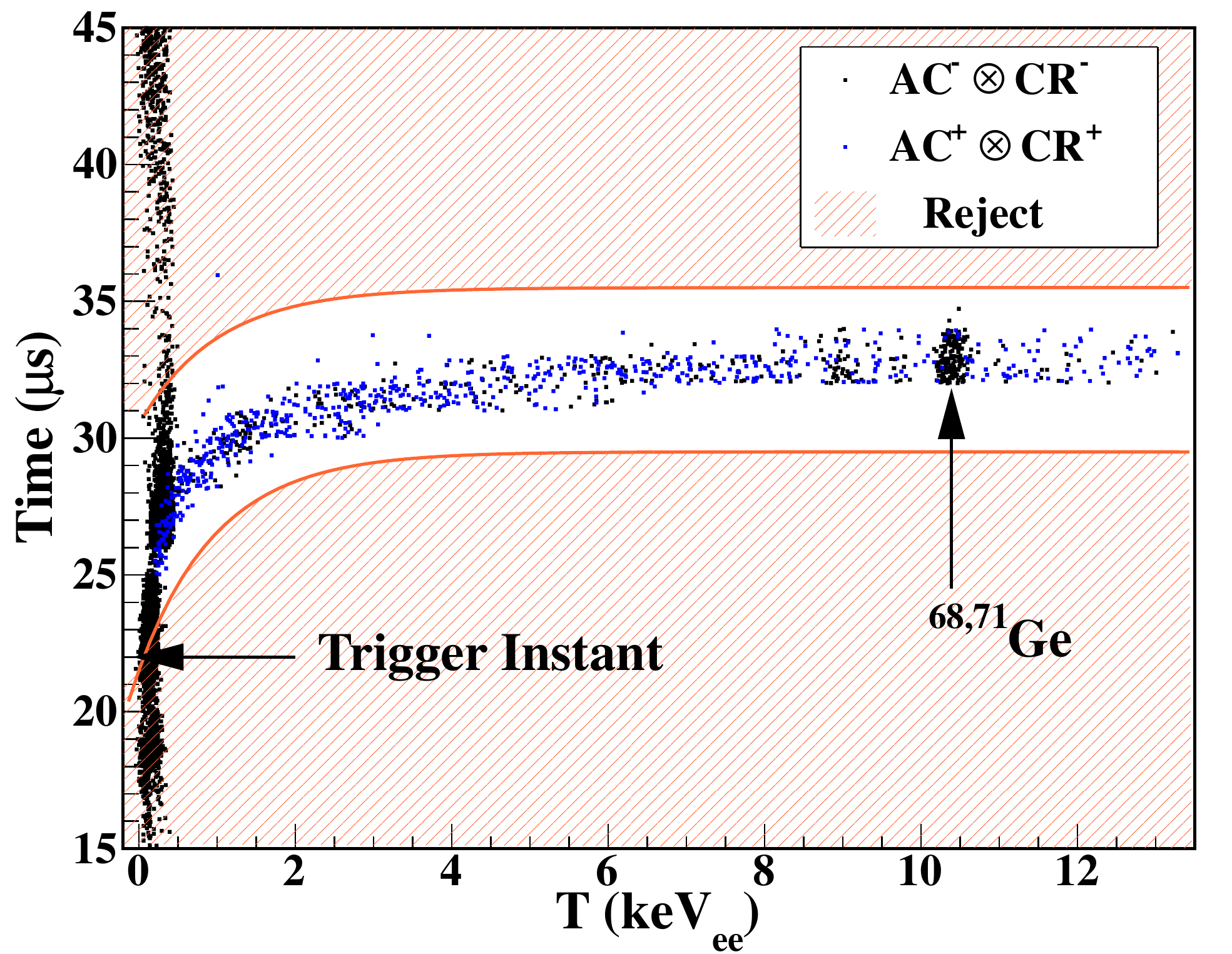}\\
\caption{
Timing correlation of maximal pulse location
(``$t_{max}$'' of Figure~\ref{fig::pulseshape}c)
relative to trigger instant.
Data from nPCGe are used as illustrations.
}
\label{fig::triggertiming}
\end{figure}


\begin{figure}
{\bf (a)}\\
\includegraphics[width=8cm]{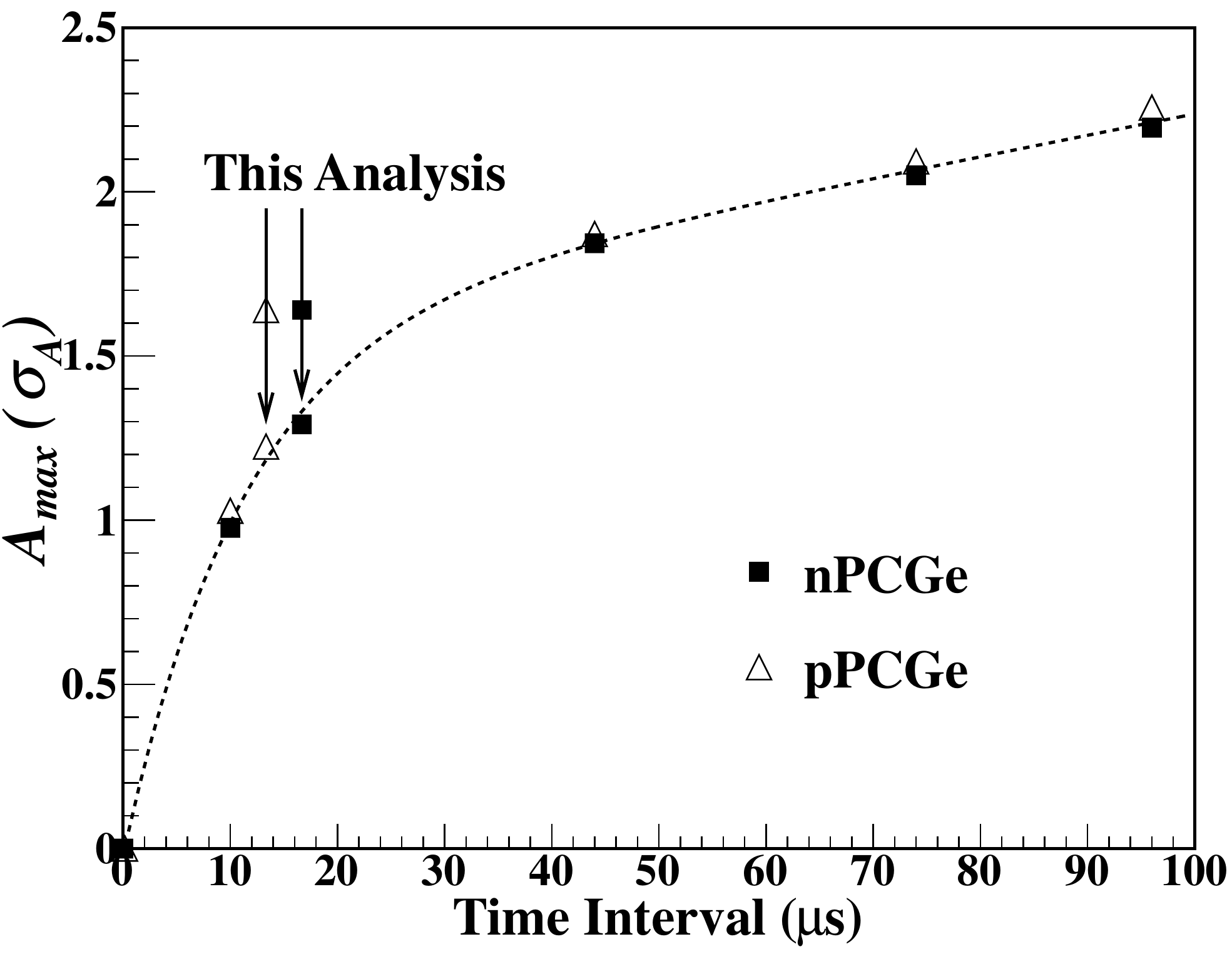}\\
{\bf (b)}\\
\includegraphics[width=8cm]{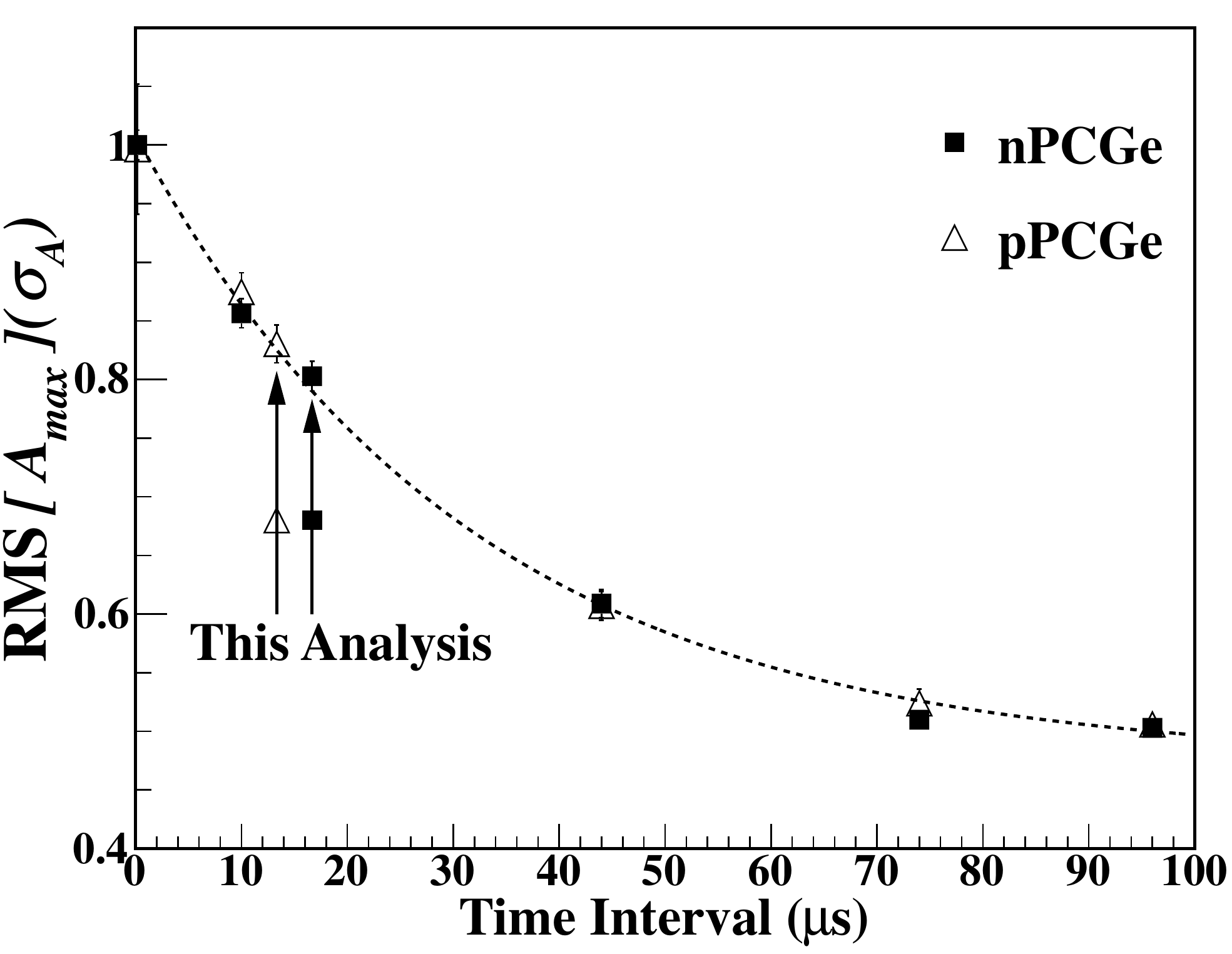}
\caption{
(a) Pulse amplitude 
$\eamp$ and (b) RMS distributions of 
random pedestal events 
as function of search time intervals after
trigger instant during which the
maximal amplitude is located.
The dotted lines are fits of combined data 
showing universal behavior independent of detectors.
The selected intervals for the analysis in this work
are marked.
}
\label{fig::timeinterval}
\end{figure}


\subsubsection{Energy Response}
\label{sect::energyresponse}

Deviations from
linearity in $\eamp$ can be expected 
when pulse amplitude is comparable to pedestal
noise fluctuations.
This is a consequence of using  pulse amplitude as 
energy estimator $-$ 
even random sampling of
the pedestal noise would give finite values of $\eamp$.
However, as shown in Figure~\ref{fig::response}b,
the response is non-linear only at $\eamp < 6 ~ \amprms$
for both detectors.
At the region of interest above the
electronic noise-edge of 7.3(7.6)~$\amprms$
for pPCGe(nPCGe), the response is linear  
with deviations $<$4~$\eVee$,
justifying the validity of the amplitude measurement.

Energy resolution of pulser-events is a good
parameter for characterizing the
contributions of electronics system and
for comparing the performance of
different detectors, thereby
providing complementary and independent information
to $\amprms$ from pedestal noise.
The RMS resolution for test-pulser input 
is shown in Figure~\ref{fig::response}c.
The energy response deteriorates at 
$\eamp < 1.2 \amprms$,
which is below the electronic noise-edge and physics 
region of interest.
                                                              
The energy response varies with
search time intervals after the
trigger instant, in which 
$\eamp$ is located.
The mean time difference between $\eamp$
and the trigger instant is depicted in 
Figure~\ref{fig::triggertiming}.
There is a shift of the maximum timing 
at low energy, because the trigger instant
is defined by a constant-voltage discriminator
level.

The distributions of $\eamp$ and their RMS  
with random pedestal events (equivalent to
pulser events with $A_{Pulser} = 0$)
as function of search time intervals are displayed in 
Figure~\ref{fig::timeinterval}a\&b, respectively.
Shorter intervals would extend 
the linearity range to lower energy  
but deteriorate energy resolution.
In the limiting case when 
a single time-bin is taken to define the pulse maximum 
(that is, when the search time interval approaches zero), 
$\eamp = 0$ and RMS$[ \eamp ] = \amprms$ 
for random-trigger events
and the energy response is 
linear down to zero-energy, as expected.
The selected intervals for analysis in this work
are marked.

The fiducial volume of pPCGe corresponds to
the region giving rise to ``Bulk'' events 
(to be described in Section~\ref{sect::bsppcge}i)
while that of nPCGe is  
the entire active detector, 
minus the $\rm{p^+}$ surface electrode
of micron-level thickness and the $\sim$mm$^3$ volume
around the $\rm{n^+}$ point-contact.
Physics events originated at
different locations of the detector fiducial volume
exhibit the same response, as
discussed in Section~\ref{sect::bsppcge}ii.
This justifies the adoption of
physics background events 
with AC$^+$$\otimes$CR$^+$ (plus ``Bulk'' tag for pPCGe)
to characterize {\it in situ}
detector behavior and to measure efficiencies
of neutrino- and WIMP-induced signals in
AC$^-$$\otimes$CR$^-$, although their
spatial locations within the detectors are
in general not identical.

The current scheme 
of energy measurement with the amplitude of $\sasix$
is therefore applicable to the entire detector fiducial volume.
It is robust and well-behaved 
in the energy range corresponding to physics events.


\subsection{Quenching Factor}
\label{sect::qf}

Quenching factor (QF) is the ratio of ionization
energy to the nuclear recoil energy 
deposited by radiations.
In Ge detectors, ionization energy corresponds to
the amount of electron-hole pairs created.
Knowledge of QF is essential in
studies on $\nu$N and $\chi$N processes in
Ge ionization detectors, the
signatures of which are due to nuclear recoils.


\begin{figure} 
\includegraphics[width=8.5cm]{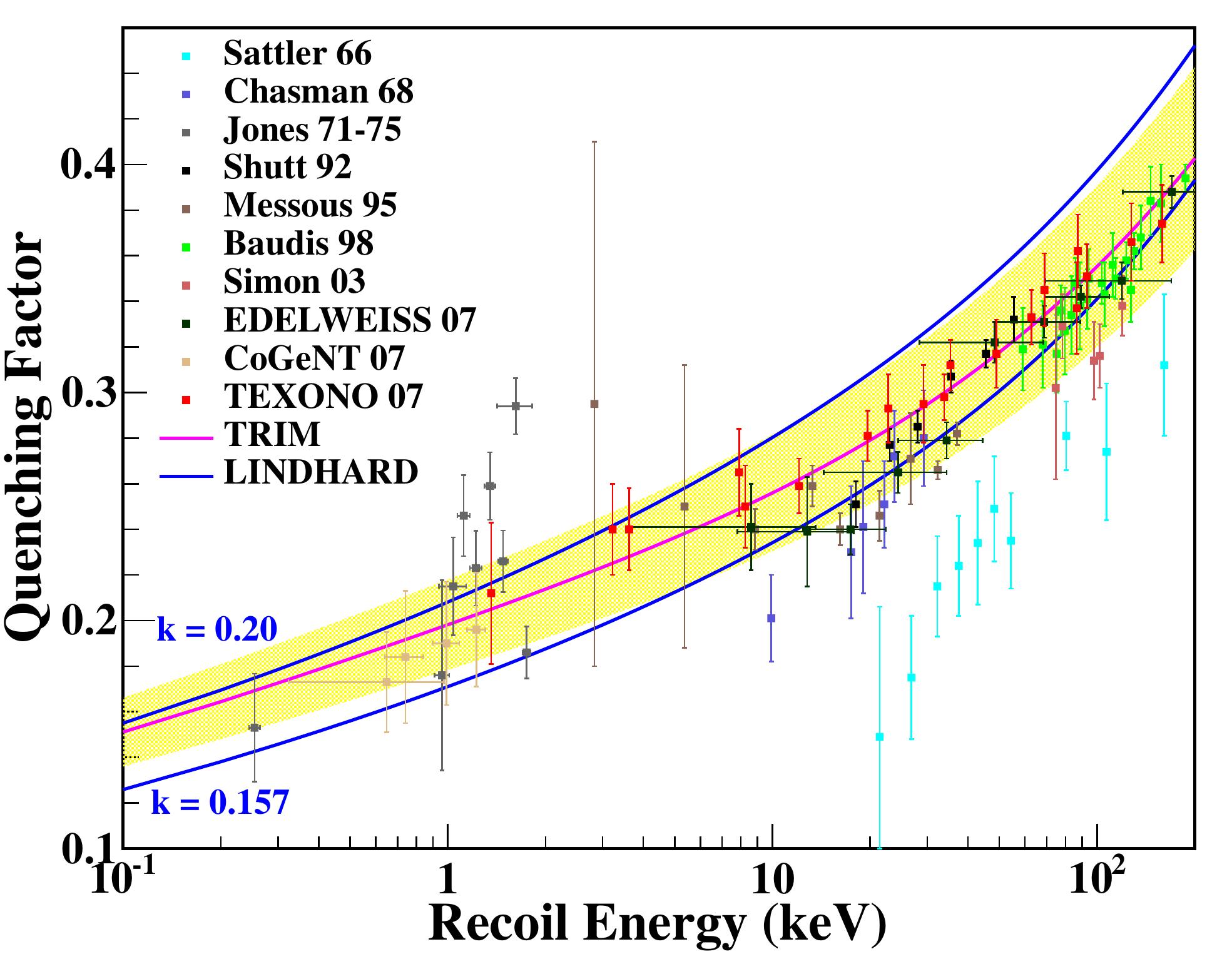}
\caption{
Summary of QF studies on Ge detectors.
Derived values obtained using 
the TRIM software package~\cite{trim} and the
Lindhard theory~\cite{lindhard} in two parametrizations 
are superimposed to the measurements.
Selection of the TRIM-QF curve with 10\% systematic
uncertainty (yellow band) provides good description
of the current knowledge over a large energy range.
Readers are advised to refer to the color version
which presents better the wealth of information.
}
\label{fig::qf}
\end{figure}


A compilation of existing
QF measurements in Ge~\cite{qfge}
is given in Figure~\ref{fig::qf}.
Superimposed are calculations obtained using
the TRIM software~\cite{trim}
and the Lindhard model~\cite{lindhard}
at two popular parametrizations (k=0.157 and 0.20).
Both schemes have been used in various dark
matter experiments.
The TRIM results explain
the QF data at a larger energy range
and were adopted in our previous analysis
with an assigned systematic uncertainty 
of 10\%~\cite{texono2009,texono2013,cdex0,cdex1}.
As illustrated by the yellow band of Figure~\ref{fig::qf},
this choice accounts well the spread of measured data
and the two Lindhard parametrizations.
Constraints on $\chi$N cross-sections
were selected from the most conservative 
results within the uncertainty band.


\section{Signal Selection and Efficiency}
\label{sect::results}

\subsection{Overview}

Signal events have to survive various selection criteria 
(``cuts'') applied to the raw data.
The survival probability or signal efficiency 
of each of these cuts must be known
in order to correctly extract the physics information.

\begin{enumerate}
\item
Trigger $-$ The physics events produce 
trigger signals for the DAQ system. 
Trigger efficiency ($\trigeff$) is energy-dependent.
Its evaluation is discussed in Section~\ref{sect::daqeff}.
\item
DAQ $-$ The trigger signal  
activates the DAQ system which 
subsequently records a complete event. 
DAQ efficiency ($\daqeff$) is energy-independent.
It can be accurately evaluated by the 
ratio of random-trigger events recorded in 
the DAQ computer to the number of
trigger signals issued.

This efficiency factor is associated with the ``DAQ dead-time'' 
$-$ the fraction of time in which the DAQ system is not actively
responding to the trigger,
through the relation
$\rm { [ DAQ~Dead~Time = 1 - \daqeff ]}$ .

\item
Analysis $-$ 
The objectives of offline analysis procedures 
are to retain physics signals and suppress background events.
Some of the signal events may be rejected in the processes,
therefore giving rise to efficiency factors which are generally 
energy-dependent.

In this work,
the offline analysis can be further classified into
three categories according to their different characteristic features
listed as follows.
\begin{enumerate}
\item 
Physics signals in Ge are selected using information
from different detector components
which are the CR and AC detectors
for KSNL measurements.
Neutrino- and WIMP-induced signals in Ge
are uncorrelated (that is, in anti-coincidence) 
with these detectors. 
The selected anti-coincidence interval
in software analysis is 4.5~$\mu$s.
The signal efficiencies ($\CReff$ and $\ACeff$)
can be accurately measured with
the survival fractions of random-trigger events
subjected to the identical selection cuts.
Measurements with experiments at 
KSNL~\cite{texonomunu,texono2009,texono2013}
give $\CReff \simeq 93 \%$ and $\ACeff > 99 \%$.
\item
Physics signals are due to genuine charge depositions. 
They can be distinguished from 
self-trigger noise events
by differences in their pulse shape and timing. 
The signal efficiencies 
can be derived from the survival probabilities of 
AC$^+$$\otimes$CR$^+$ samples.\\
We present several selection criteria 
with general applicability to all detectors
in Section~\ref{sect::basicfilters},
where their combined efficiency 
is denoted by $\bfeff$.
Advanced algorithms with goals of
extracting physics signals {\it below} the
noise-edge are being pursued.
\item
Events due to neutrino and WIMP interactions are
mostly located in the bulk of the detectors while 
background events from low-energy
ambient radioactivity deposit their energy primarily
at the surface. 
These events differ in their rise-time 
in pPCGe~\cite{gesurface}.
Selection of bulk events requires pulse-shape analysis.  
Special calibration schemes are devised to evaluate
the signal efficiency and background contamination 
factors~\cite{bsel2014}.
\end{enumerate}
\end{enumerate}

We emphasize that, in order to avoid bias,
a single energy measurement scheme
($\eamp$ as defined in Figure~\ref{fig::pulseshape}c) 
should be used exclusively and consistently 
throughout the
sequence of selection cuts and efficiency evaluation.


\begin{figure}
{\bf (a)}\\
\includegraphics[width=8.2cm]{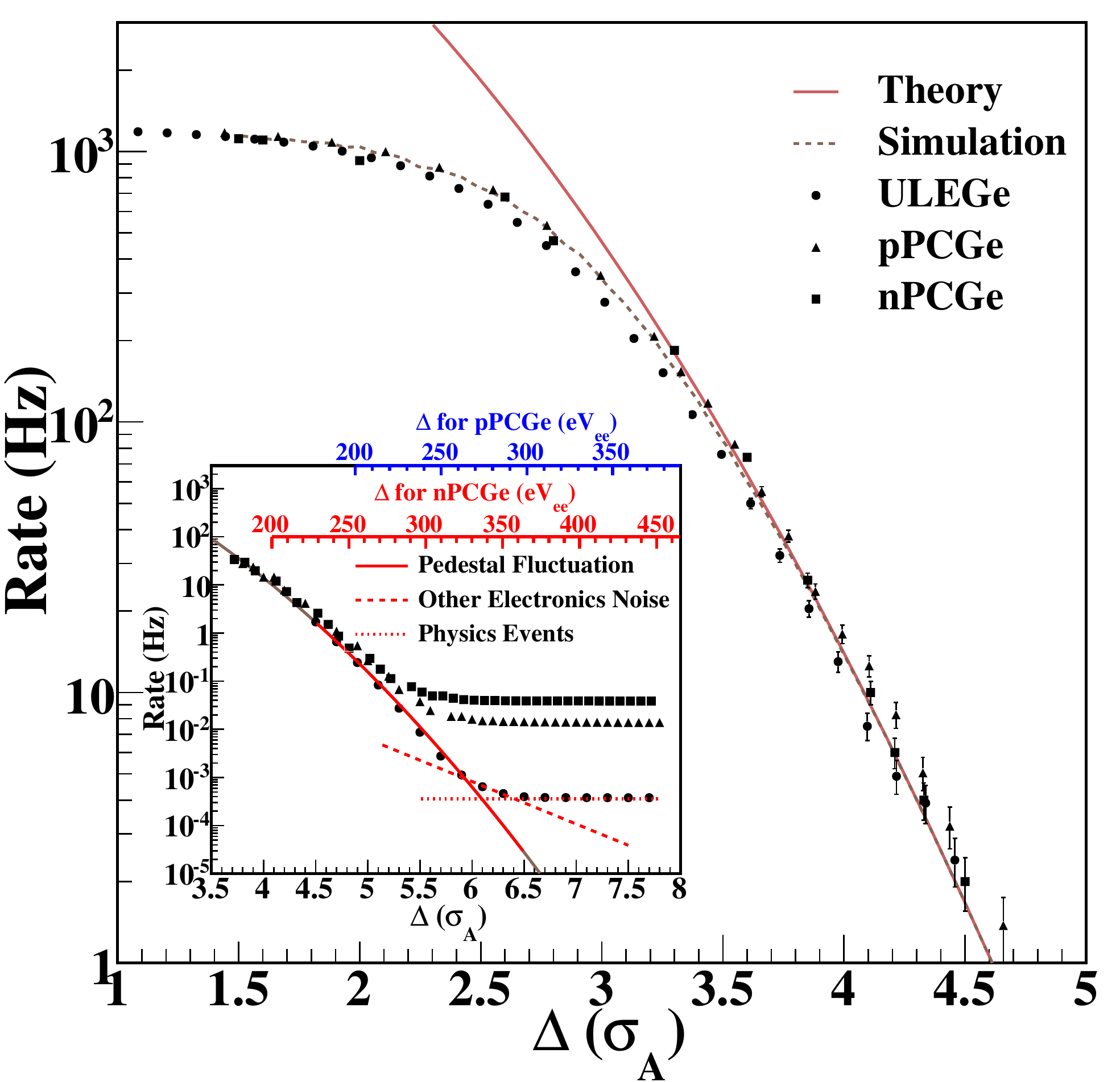}\\
{\bf (b)}\\
\includegraphics[width=8.2cm]{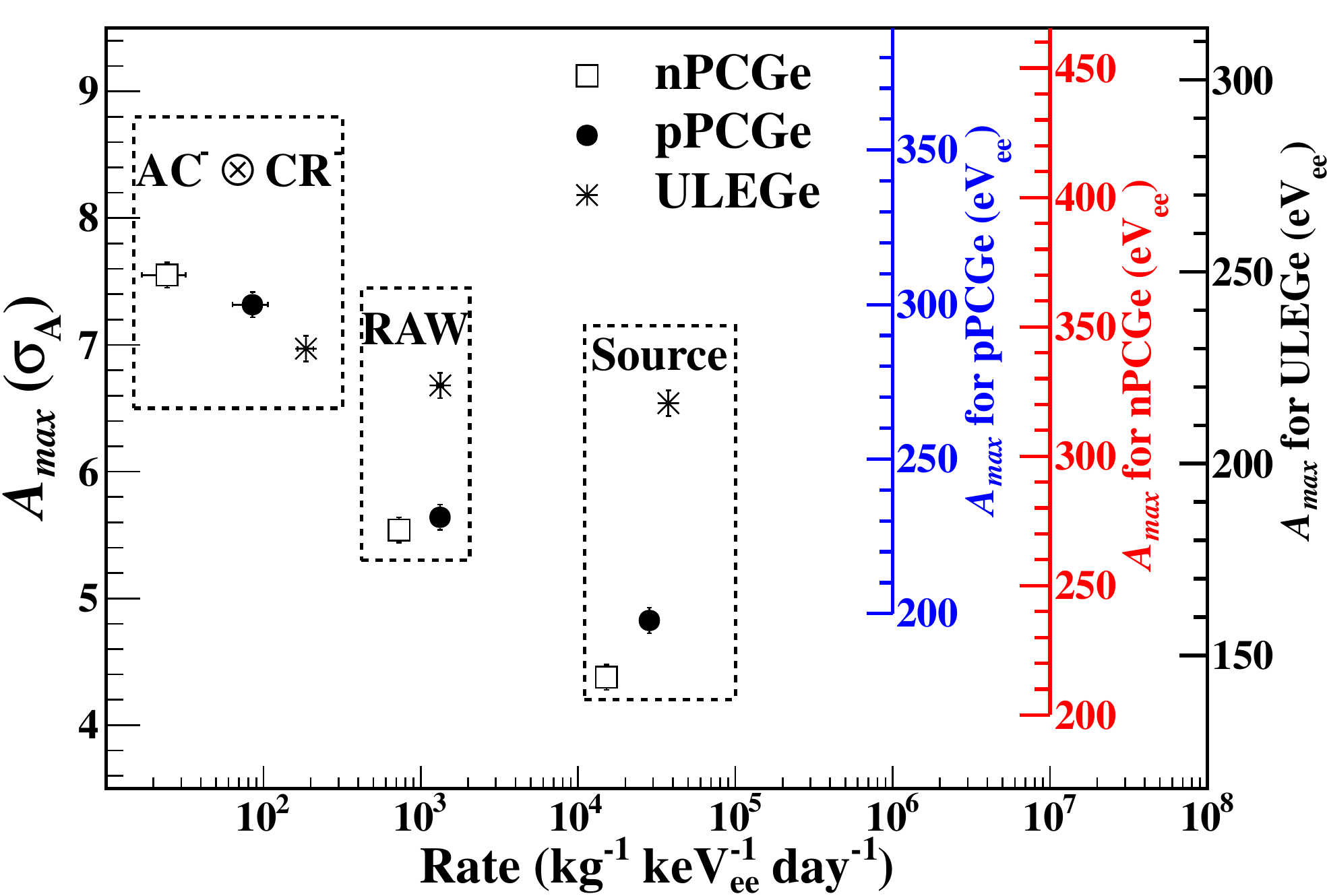}\\
{\bf (c)}\\
\includegraphics[width=8.2cm]{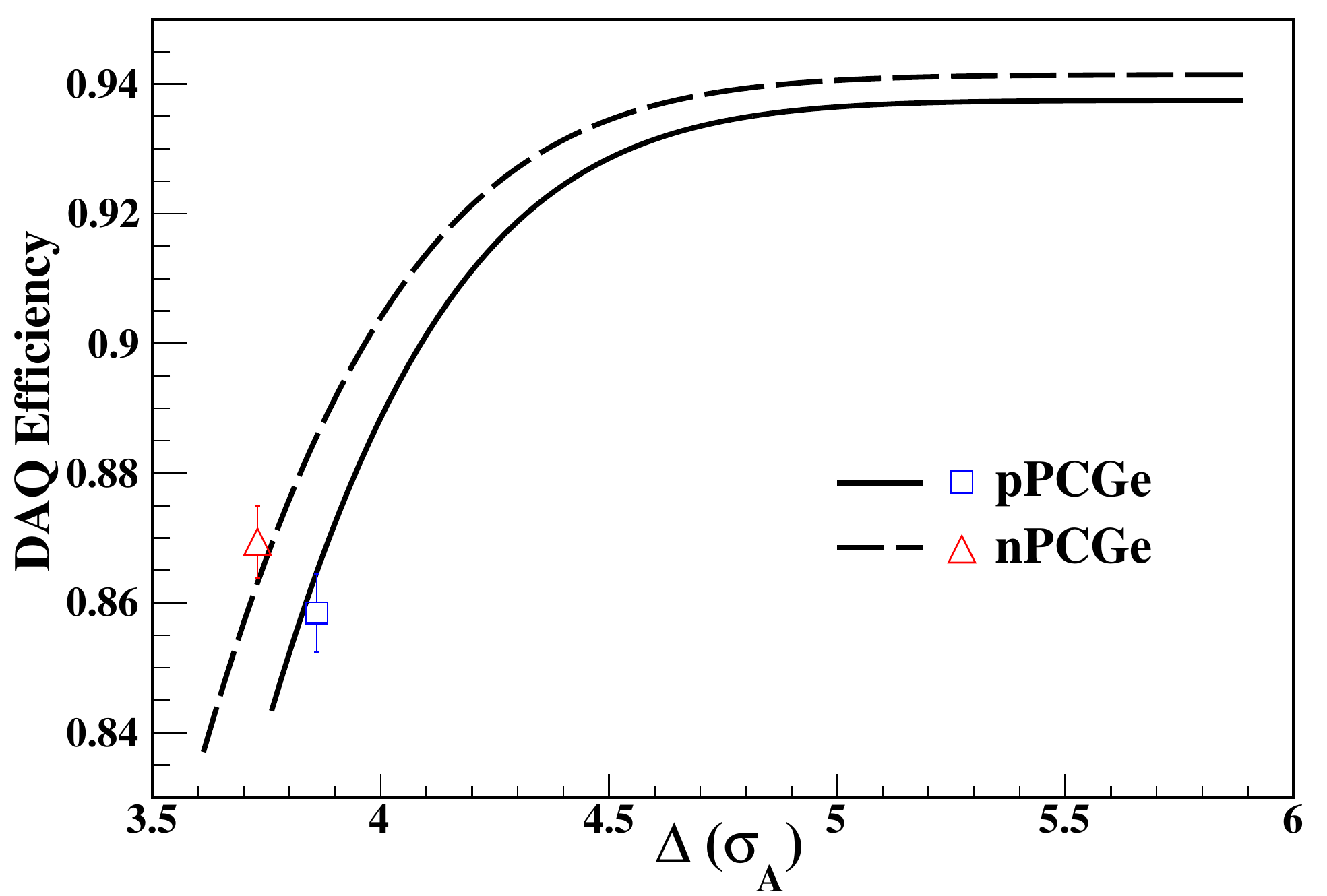}
\caption{
(a) 
Measured trigger rates as a function of discriminator
threshold $\Delta$, showing a universal behavior 
of self-trigger pedestal noise when
expressed in $\amprms$ unit. 
The origins of deviations at high and low
$\Delta$ are explained in the text.
The energy scales for the specific detectors
in $\eVee$ are also shown.
(b)
Noise-edge versus rates of physics event
for various detectors and at different
background configurations: (i) high rates with
radioactive sources, (ii) raw rates at
low-background configurations at KSNL, and
(iii) after AC$^-$$\otimes$CR$^-$ selection.
The data demonstrate
that lower event rates
correspond to higher noise-edge levels
for the same detectors.
(c)
Data acquisition efficiencies $-$
the expectation follows directly from (a) 
using as input the average DAQ dead-time per event 
and the post-RESET veto windows.
Measurements are provided by random-trigger events.
}
\label{fig::trate}
\end{figure}


\subsection{Trigger and Data Acquisition}
\label{sect::trig}

\subsubsection{Trigger Rates}
\label{sect::trig-rate}

Trigger instant is defined as the time 
at which $\sasix$ amplitude $\eamp$ 
exceeds the discriminator level $\Delta$.
When the amplitude of pedestal fluctuations
follows a Gaussian distribution, 
the trigger rate is given by~\cite{triggerrate}:
\begin{equation}
\label{eq::daqrate} 
R \sim \frac{1}{4 ~ \tau_S} ~ 
{\rm exp}  ~ [ - \frac{1}{2} ~ ( \frac{\Delta}{\amprms} )^2  ~ ]  ~~,
\end{equation}
where $\tau_S$ is the shaping time and 
$\amprms$ is the RMS of pedestal fluctuation.
Measurements of random-trigger events
show there is on average one positive fluctuation
per 4 $\tau_S$, verifying 
the normalization of Equation~\ref{eq::daqrate}.


The measured trigger rates as a function of $\Delta$ 
are displayed in Figure~\ref{fig::trate}a, showing 
consistency with predictions
at $\Delta > 3.3 ~ \amprms$
and implying that the pedestal noise
fluctuations are indeed Gaussian.
Deviations at small $\Delta$  
can be quantitatively accounted for by the 
finite width of inactive interval 
after a valid trigger, 
such that accidental triggers within the gate interval
are counted as a single event.  
In addition, all the detectors exhibit 
consistent behavior,
indicating that Equation~\ref{eq::daqrate} is universal when
the energy scale is expressed in $\amprms$ unit.
As illustrated in the inset of Figure~\ref{fig::trate}a,
various modes of ``other electronics noise'' 
(such as microphonics and RESET-induced events)
are the main contributors at low energy above
pedestal noise fluctuations, 
while physics events would dominate 
the trigger rates at large $\Delta$.

The evaluation of electronic noise-edge values
from measured energy spectra is illustrated
in the insets of Figures~\ref{fig::spectra}a,b\&c.
The self-trigger noise events are due to 
tails in Gaussian fluctuations. 
The noise-edge corresponds to the energy
above which physics events dominate
over electronic noise.
Accordingly, the noise-edge is not a constant
but varies with physics event rates.
This is depicted in Figure~\ref{fig::trate}b
for the various detectors at different
event rates due to different 
background configurations.
The measurements demonstrate
that, in a given detector,
lower physics event rates
give rise to higher noise-edge levels.
The noise-edge at KSNL
after AC$^-$$\otimes$CR$^-$ selection
corresponds to $\sim 7 ~ \amprms$. 

In this study, the $\sasix$ 
amplitude $\eamp$ is the relevant quantity
for defining online trigger and for providing
energy measurement. 
Advances in real-time pulse-shape processing
in Field Programmable Gate Arrays (FPGA)
make it possible to have other sophisticated 
and more efficient trigger schemes. 
This line of research is being pursued, with the goals 
of improving the DAQ dead time and reducing 
computing overhead. The measured detector 
performance parameters reported in this article, however, 
are independent of trigger configurations
and DAQ systems being adopted.


\begin{figure}
\includegraphics[width=8.5cm]{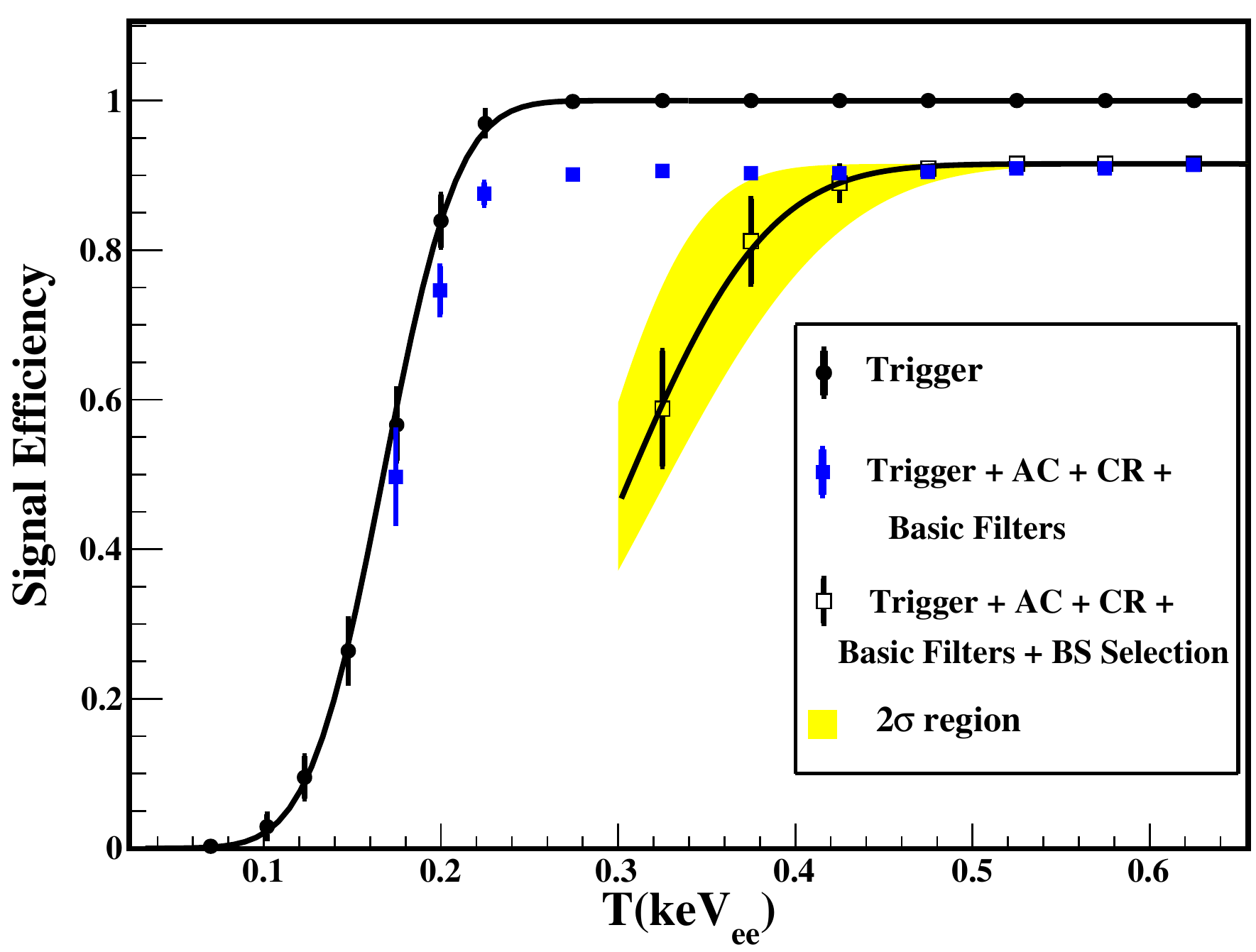}
\caption{
Various analysis efficiency factors for pPCGe
in this work.
Efficiencies before and after bulk-event
BS-selection are shown.
The signal efficiency for nPCGe is 
a step function smeared by energy resolution
at trigger level of 204~$\eVee$.
The displayed data do not include
readout efficiency factors due to DAQ dead time.
}
\label{fig::combinedeff}
\end{figure}


\subsubsection{Trigger and Data Acquisition Efficiencies}
\label{sect::daqeff}

In this analysis where both trigger timing 
and energy measurements are defined by the amplitude
of the $\sasix$ pulse, $\trigeff$ is 
a step function at $\Delta$ smeared by energy resolution.
Measurements are provided by the test-pulser data, where $\trigeff$ 
is the fraction of events giving trigger signals.
The measurements are depicted in Figure~\ref{fig::combinedeff} 
for pPCGe.
In practice, trigger rates are set to 1$-$10~Hz
corresponding to $\Delta \sim 4.6 - 4.1 ~ \amprms$, respectively. 
This $\Delta$-range is lower than the 
typical electronic noise-edge at 
$\sim 5 - 6 ~ \amprms$, such that $\trigeff$ 
does not play a role in the physics analysis.

The DAQ dead-time corresponds to two circumstances:
(i) in a certain pre-set time window (typically 10~ms)
after the preamplifier RESET, and
(ii) during data transfer to the computer hard disk 
(typically 2~ms per event).

The expected evolution of $\daqeff$ 
with $\Delta$ in Figure~\ref{fig::trate}c
follows directly the measured trigger rates of 
Figure~\ref{fig::trate}a, 
using as input 
the average DAQ dead-time per event
and the post-RESET veto windows.
Direct measurements with
ratios of recorded events to number
of random-triggers issued to the DAQ system
are superimposed, showing excellent agreement.
At large $\Delta$, $\daqeff$ is constant
and defined by the post-RESET veto time.
Variations in the constant level
among detectors can be attributed to
differences in RESET time intervals.


\subsection{Signal Events Selection} 

The exact selection procedures and efficiency derivations
for physics signals,
though differ in details for each detector and in each analysis,
share many common features.


\begin{figure}
{\bf (a)}\\
\includegraphics[width=8.2cm]{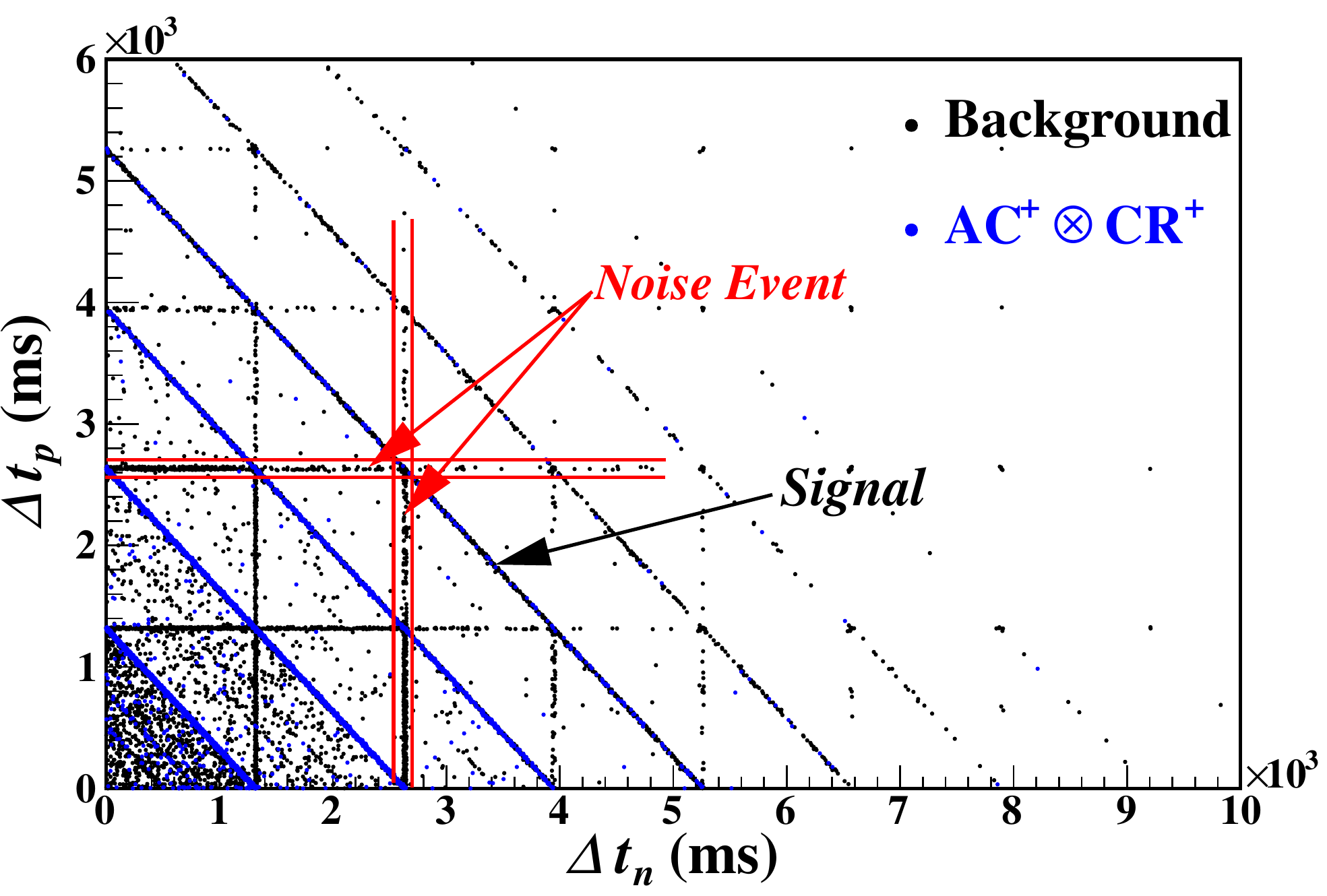}\\
{\bf (b)}\\
\includegraphics[width=8.2cm]{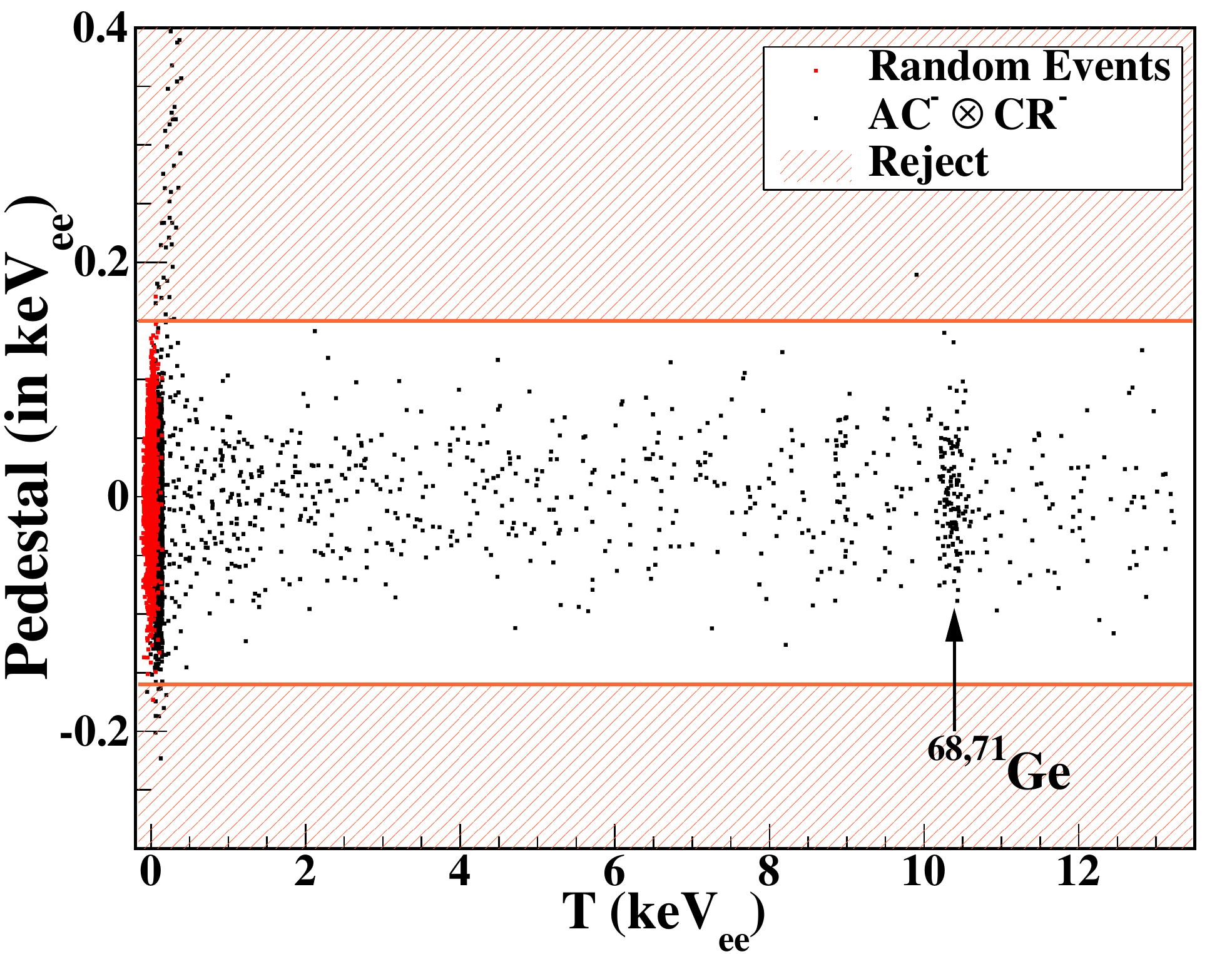}\\
\caption{
Basic filter selections, provided by
(a) timing within a RESET period
($\Delta t^+$ and $\Delta t^-$ of Figure~\ref{fig::pulseshape}b),
and
(b) pedestal fluctuations. 
An additional selection is on the timing of $\sasix$ maxima, 
displayed in Figure~\ref{fig::triggertiming}.
``Reject'' parameter spaces are shown as shaded regions.
Data from pPCGe and nPCGe are used as illustrations
in (a) and (b), respectively.
}
\label{fig::basicfilter}
\end{figure}


\subsubsection{Basic Filters}
\label{sect::basicfilters}

RESETs from preamplifiers induce
electronic noise which gives rise to
self-trigger events.
As depicted in Figure~\ref{fig::basicfilter}a, 
these background events can be identified 
via correlations in  
timing with the previous and next 
RESETs ($\Delta t^-$ and $\Delta t^+$, respectively,
as defined in Figure~\ref{fig::pulseshape}b).
There is a natural RESET period for every detector
(which is $\sim$1.3~s in this example of pPCGe
with an earlier preamplifier 
model\footnote{Model PSC921, Canberra Lingolsheim.}).
Most physics signals populate the dominant diagonal bands 
at low-($\Delta t^+ , \Delta t^-$).
Events at those longer-period bands 
correspond to situations where  
the previous or next RESETs are not recorded when  
they fall in the DAQ dead-time intervals.
Some RESETs would be issued prior to their natural period
when they are induced by electronics instabilities or 
when the detector is saturated by large charge depositions,
such as direct hits of cosmic-ray muons.
RESET-induced noise events are the horizontal and vertical
bands on the plot, and can be completely rejected.

Other electronic noise events include 
those induced by the tails of earlier signals. 
Characterized by 
an anomalous pedestal level prior to the trigger instant,
these noise events can be efficiently 
identified and rejected,
as demonstrated in Figure~\ref{fig::basicfilter}b.

The timing of the $\sasix$ maxima,
depicted in Figure~\ref{fig::triggertiming},
is another effective selection criterion. 

The corresponding signal efficiencies 
$\bfeff$ are derived.
For the $\Delta t^+ {\rm -} \Delta t^-$ selection 
of Figure~\ref{fig::basicfilter}a,
$\bfeff > 97\%$,
given by the survival fraction of 
AC$^+$$\otimes$CR$^+$ physics events
when subjected to identical criteria. 
The selection of pedestal range illustrated in
Figure~\ref{fig::basicfilter}b has $\bfeff > 99\%$,
given by random-trigger events. 
The selection of pulse maxima
shown in Figure~\ref{fig::triggertiming} 
has $\bfeff > 99\%$,
derived also with AC$^+$$\otimes$CR$^+$ physics events.
These calibration events 
are also displayed in the respective figures.


\subsubsection{Pulse-Shape Analysis on Event Rise-Time}
\label{sect::bsppcge}

Electron-hole pairs produced at the
surface (S) layer in pPCGe
are subjected to a weaker drift field
than those in the bulk volume (B).
A fraction of the pairs will recombine while the
residuals will induce signals which are weaker
and slower than those originated in B.
The S-events would therefore exhibit slower rise-time and
partial charge collection compared to
B-events~\cite{gesurface}.
This effect becomes relevant to 
the selection of neutrino- and 
WIMP-induced signals at sub-keV energy.
Quantitative studies on these detector features,
the differentiation of S and B events by pulse-shape
analysis and the calibration schemes which provide
measurements of signal-retaining and 
background-suppression
efficiencies ($\effbs , \lmbdbs$) 
are discussed in
detail in Ref.~\cite{bsel2014}.
We present new results on both pPCGe and nPCGe
in what follows.

\begin{description}
\item[(i)]{\bf Bulk versus Surface Events in pPCGe} 


\begin{figure}
{\bf (a)}\\
\includegraphics[width=8.2cm]{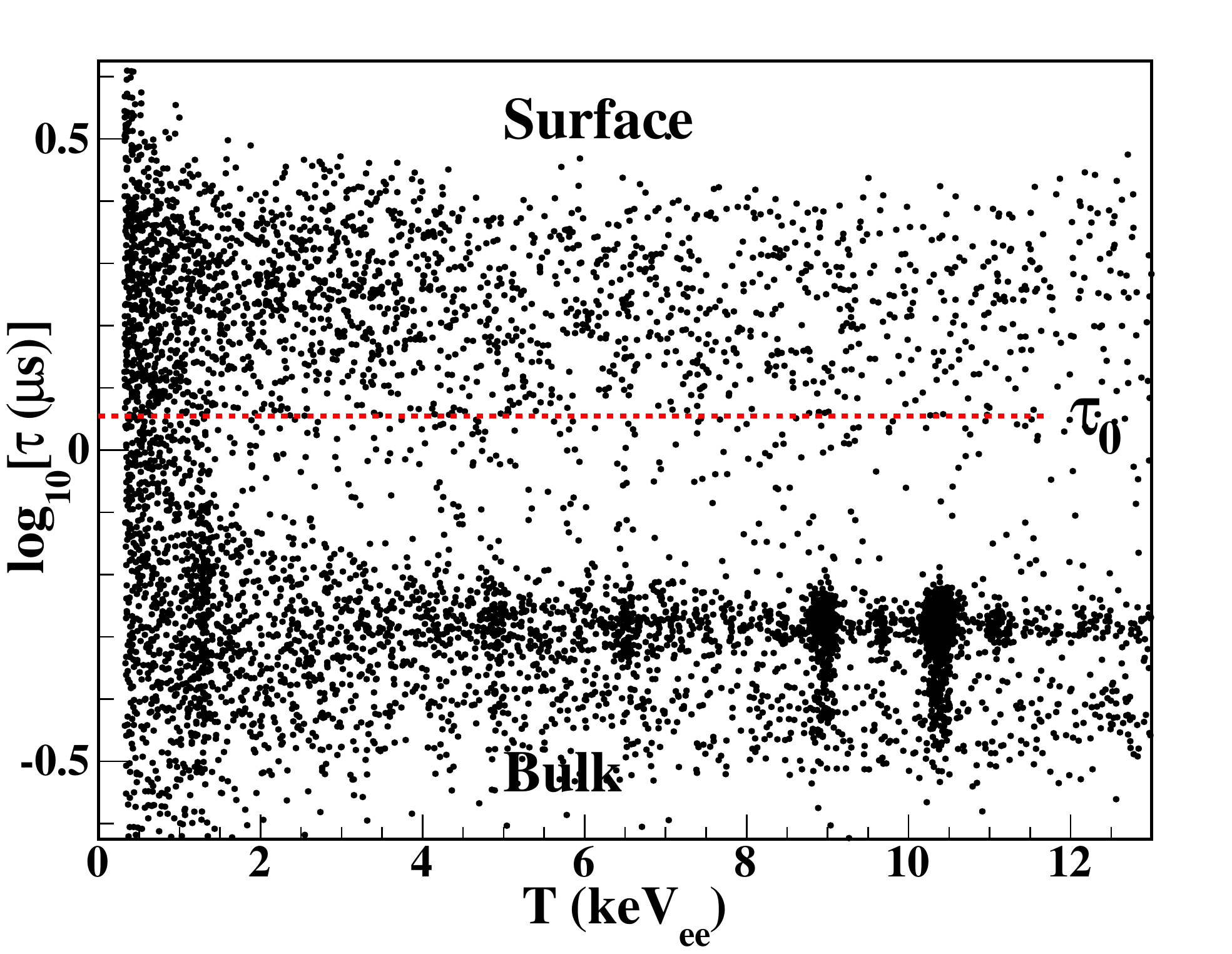}\\
{\bf (b)}\\
\includegraphics[width=8.2cm]{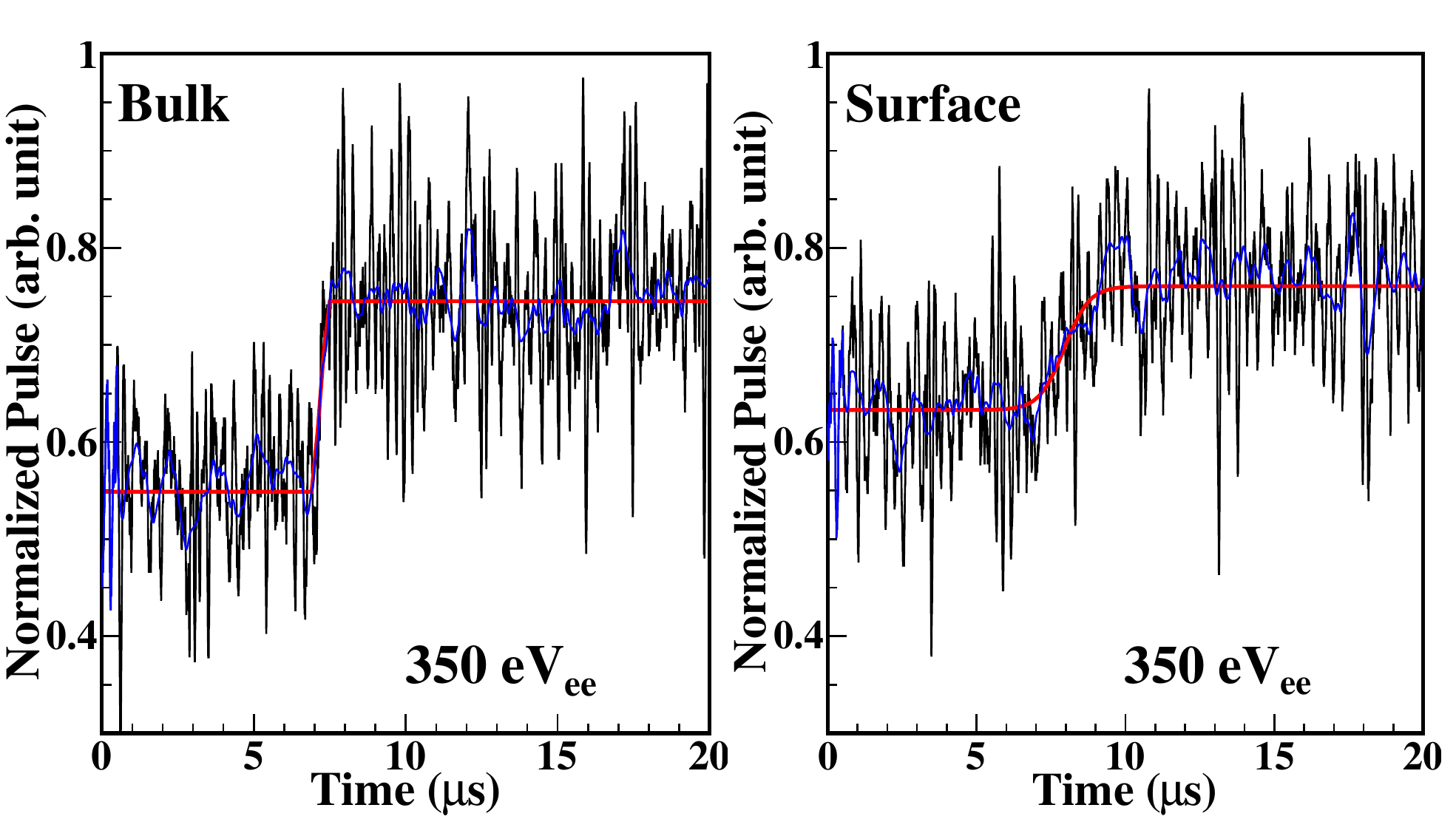}
\caption{
(a)
TA $\tau$-distribution in
pPCGe with low background at KSNL.
The selection criterion of B signal events
is defined by the cut at $\tau_0$.
(b)
Typical TA pulses for B and S events near threshold
for pPCGe.
}
\label{fig::bstau-ppc}
\end{figure}

\begin{figure}
{\bf (a)}\\
\includegraphics[width=8.2cm]{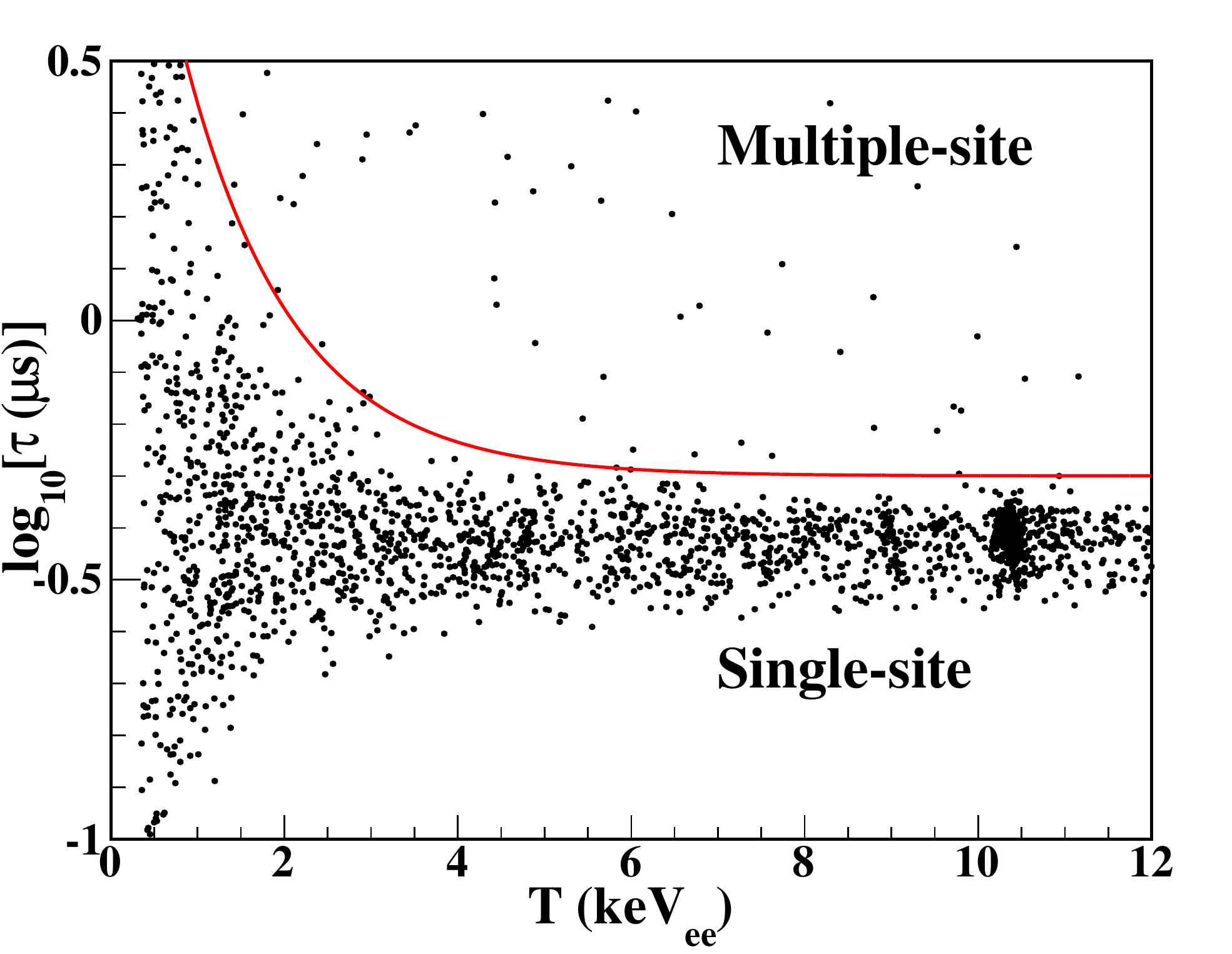}\\
{\bf (b)}\\
\includegraphics[width=8.2cm]{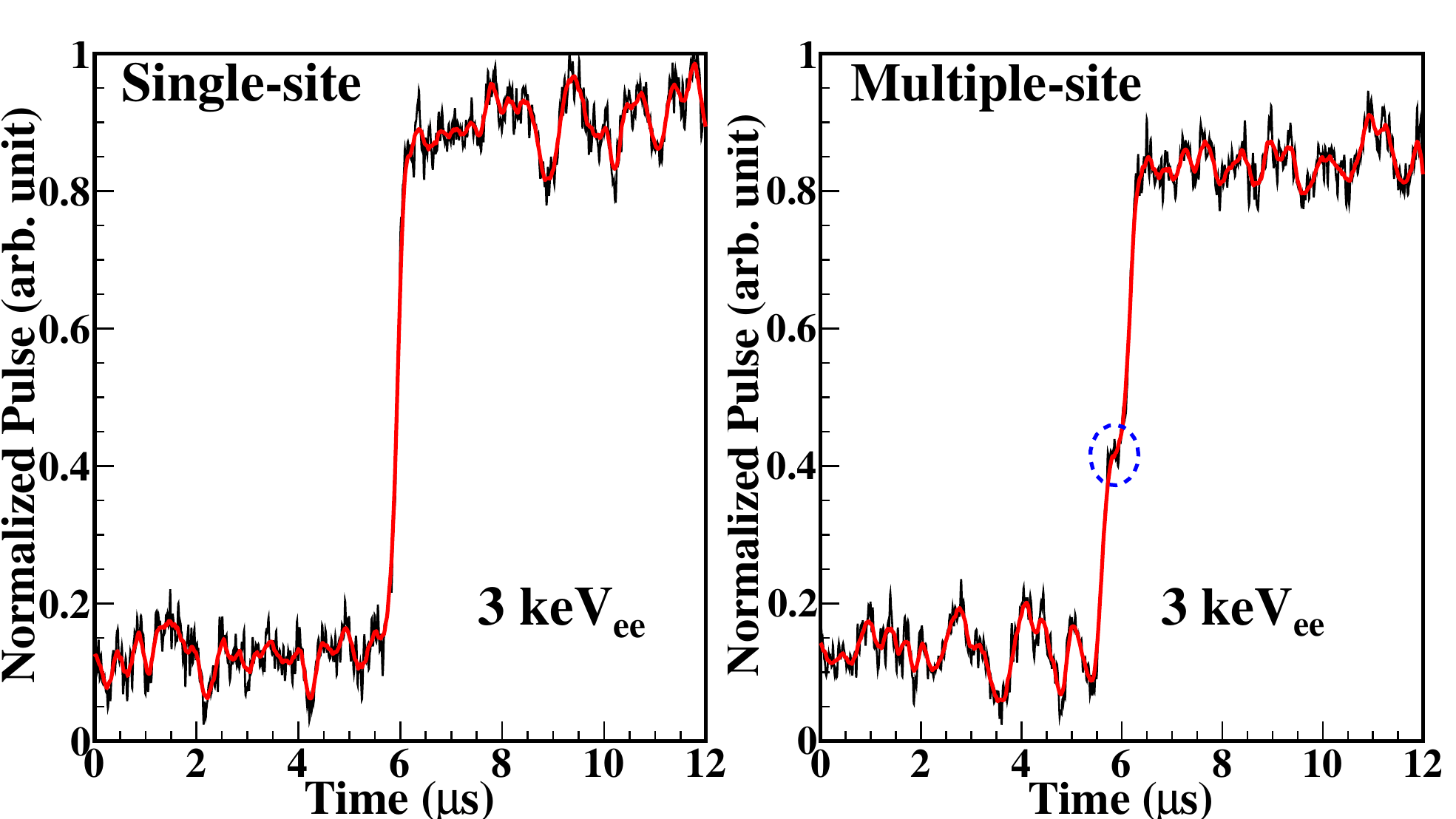}
\caption{
(a) 
TA $\tau$-distribution in nPCGe with 
low-background at KSNL, 
showing a dominating band due to
single-site events in the detector bulk.
Scattered ``slow-$\tau$'' events
away from the band are from 
multiple-site energy depositions.
Anomalous surface events
are negligible.
(b)
Typical TA pulses showing single-
and multiple-site events
for nPCGe.
Multiple-site events 
are characterized by kinks in rise-time profiles
like the one marked in blue circle.
}
\label{fig::bstau-npc}
\end{figure}



\begin{figure}
{\bf (a)}\\
\includegraphics[width=8.5cm]{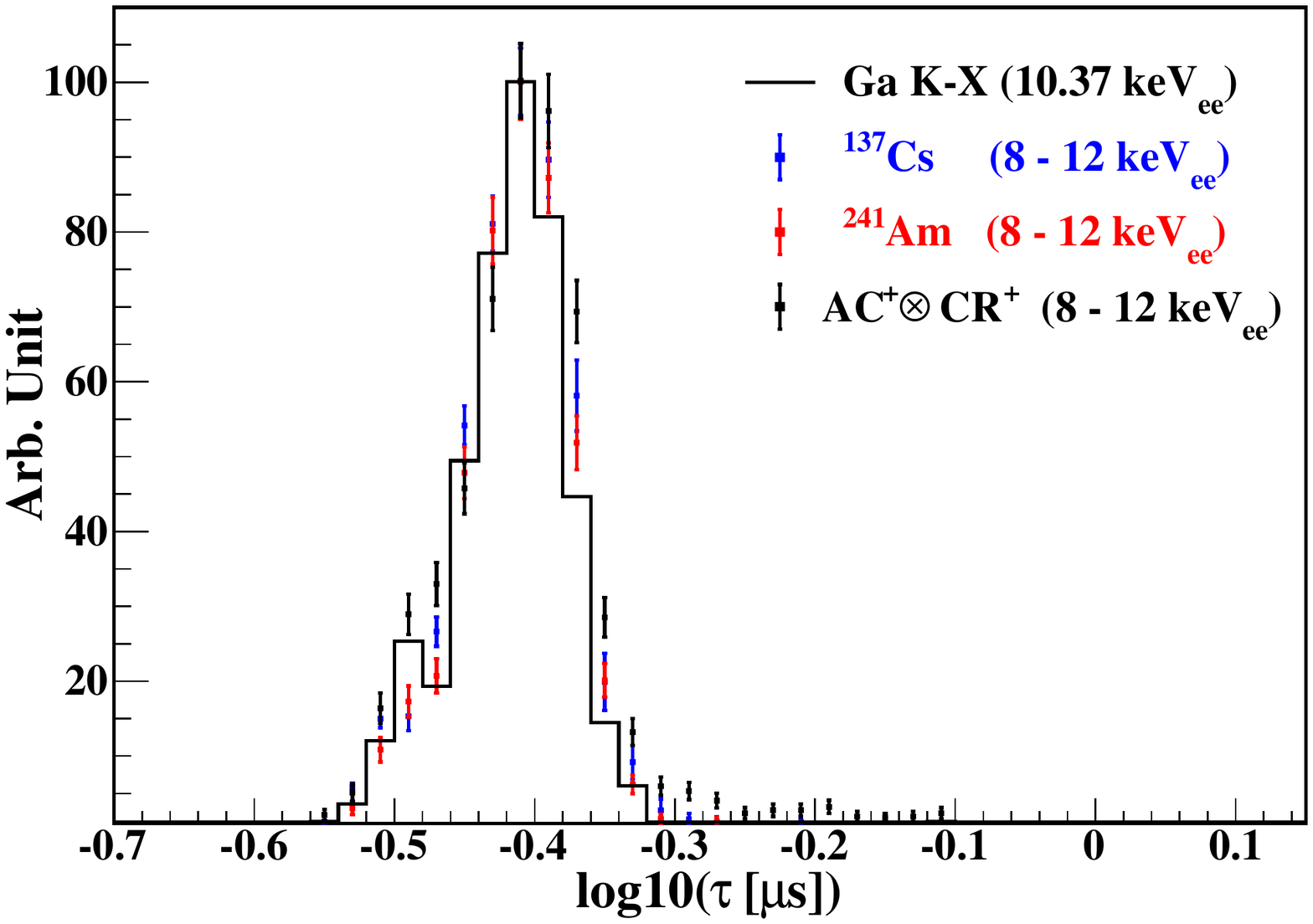}\\
{\bf (b)}\\
\includegraphics[width=8.5cm]{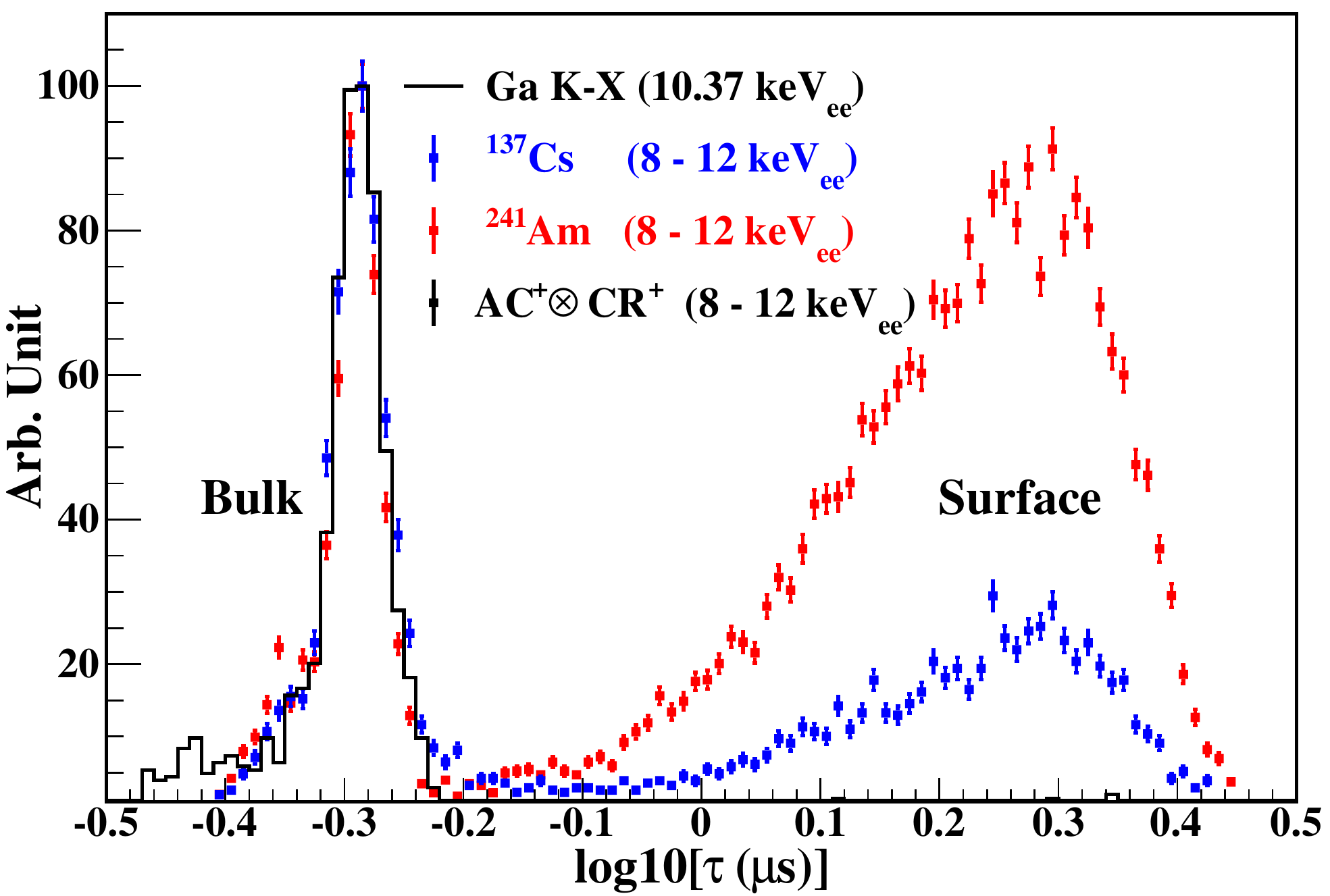}
\caption{
TA $\tau$-distributions of
$^{241}$Am, $^{137}$Cs, AC$^+$$\otimes$CR$^+$, 
and Ga K-shell X-rays
events in (a) nPCGe and (b) pPCGe.
There are no anomalous surface events in nPCGe.
Bulk-to-Surface event ratios in pPCGe are different,
because different sources give rise to events with
different spatial distributions in the detector.
Bulk distributions for all sources are consistent 
in both detectors, demonstrating uniform timing 
response over the entire detector fiducial volume.
}
\label{fig::taudist}
\end{figure}


\begin{figure}
{\bf (a)}\\
\includegraphics[width=8.5cm]{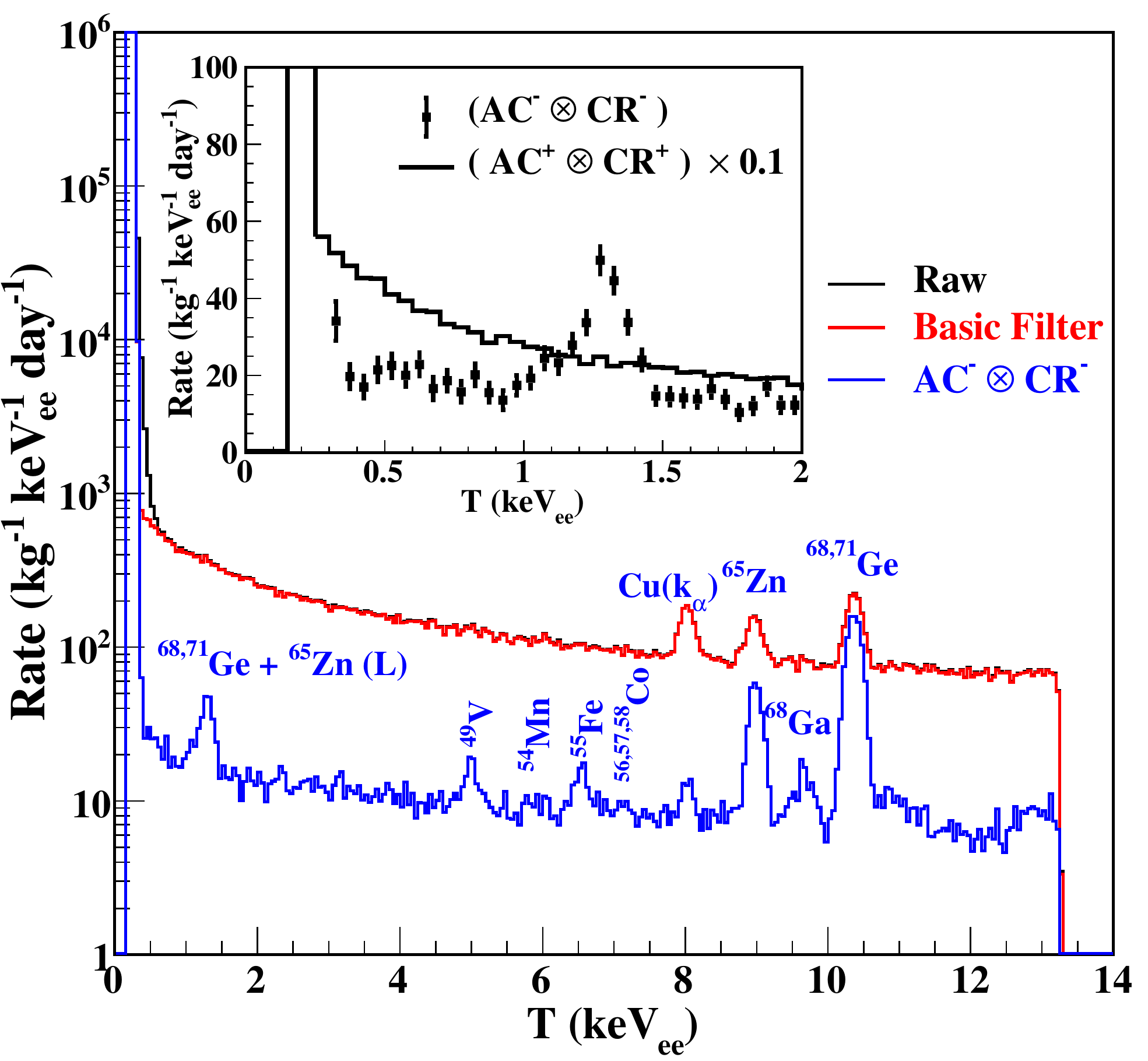}\\
{\bf (b)}\\
\includegraphics[width=8.5cm]{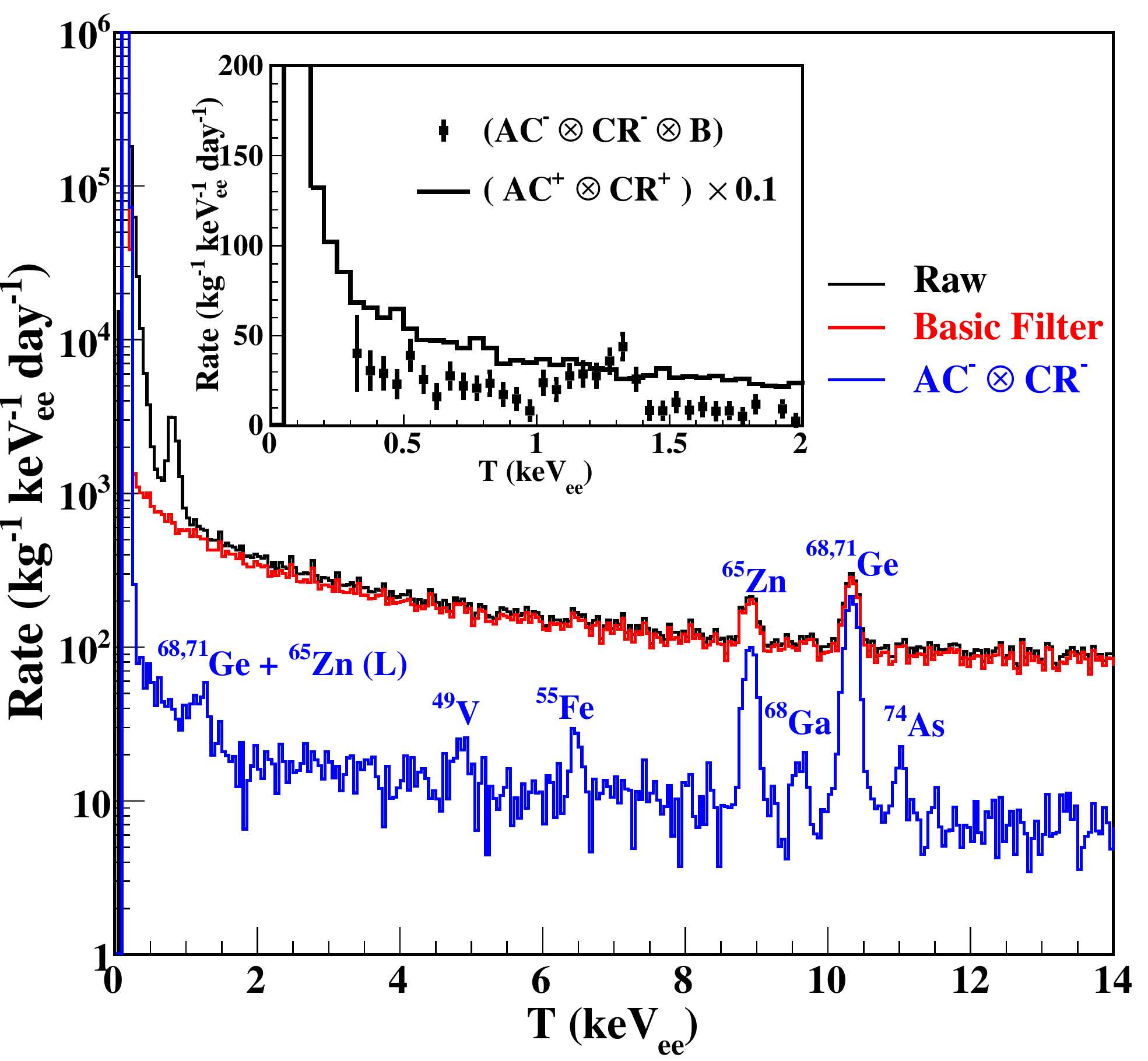}
\caption{
Evolution of AC$^-$$\otimes$CR$^-$ spectra 
taken with (a) nPCGe and (b) pPCGe
at KSNL, from RAW via
basic filters of Figures~\ref{fig::basicfilter}a,b\&c,
AC and CR vetos, and the BS-selection for pPCGe.  
The triple-coincident AC$^+$$\otimes$CR$^+$ samples
are superimposed in the insets, 
demonstrating that physics events
can be probed to a threshold below 
the electronic noise-edge.
}
\label{fig::pcspect}
\end{figure}


The rise-time of the TA-signal ($\tau$) is parametrized by
the hyperbolic tangent function
\begin{equation}
\frac{1}{2}~{\rm A_0}
\times ~ \tanh(\frac{t-{\rm t_0}}{\tau}) ~ + ~ {\rm P_0} ~~ ,
\label{eq::taufct}
\end{equation}
where $\rm{A_0}$, ${\rm P_0}$ and $\rm{t_0}$ are
the amplitude, pedestal offset and timing offset, respectively.
The $\tau$-distribution of pPCGe events 
with AC$^-$$\otimes$CR$^-$ tags 
is displayed in Figure~\ref{fig::bstau-ppc}a.
At high energy where electronic noise is negligible, 
the fits are in excellent agreement with data
(with a mean $\chi ^2$/dof of 64.7/64),
showing that Equation~\ref{eq::taufct}
is an appropriate description of the rise-time
of physics events.

Typical B and S events near noise-edge
are illustrated in Figure~\ref{fig::bstau-ppc}b.
At low energy, the cross-contaminations between the
two samples must be taken into account.
The efficiency-corrected bulk and surface rates
($\b0 , \s0$) are related to measured rates (B, S) via:
\begin{eqnarray}
\b0 & = & 
\frac{ \lmbdbs \cdot {\rm B } - ( 1 - \lmbdbs ) \cdot {\rm S } }
{ ( \effbs + \lmbdbs - 1 ) } \nonumber \\
\s0 & = & 
\frac{ \effbs \cdot {\rm S } - ( 1 - \effbs ) \cdot {\rm B } }
{ ( \effbs + \lmbdbs - 1 ) } ~~ . 
\label{eq::b0s0}
\end{eqnarray}
Two components contribute to $\b0$($\s0$).
The first positive term accounts for 
the loss of efficiency in the measurement of B(S),
while the second negative term corrects  
mis-identification due to leakage effects.
Both ($\effbs , \lmbdbs$) factors 
should be properly accounted for in
order to provide correct measurements of the 
energy spectra for bulk events.

Systematic uncertainties in the derivation of
($\effbs , \lmbdbs$) originate from
the choice of parameters such as the cut-value $\tau_0$,
and from differences in detector response between 
the calibration sources and physics background.
However, the combined systematic and statistical
errors of ($\effbs , \lmbdbs$),
as well as the efficiencies of other cuts, 
play only minor roles in the overall uncertainties.
In low count-rate experiments, 
the leading contributions remain the statistical errors in
physics measurements~\cite{texono2013,bsel2014,cdex1}.
These are further boosted by the 1/$( \effbs + \lmbdbs - 1 )$
factor of Equation~\ref{eq::b0s0} as 
($\effbs , \lmbdbs$) deviate from unity at low energy.

In contrast, anomalous surface events in nPCGe
are negligible, 
as indicated in Figure~\ref{fig::bstau-npc}a
where the S-band is absent.
Scattered events on the ``slow-$\tau$'' side
away from the B-events are due to
charge depositions in multiple sites. Typical single-
and multiple-site events at low energy are illustrated
in Figure~\ref{fig::bstau-npc}b.

For completeness,
we note that pulse-shape analysis on
fast TA signals in pPCGe is 
also an important technique
in double beta decay experiments~\cite{dbdpsd}.
Single- and multiple-site events at the MeV range
are distinguished by comparing amplitude and energy 
of the fast TA pulse.


\input{table_summary_selection}


\item[(ii)]{\bf Uniformity within Fiducial Volume}

Signal events originated at
different parts of the detector fiducial volume
exhibit the same pulse shapes for both TA and $\sasix$.
This feature
can be verified by comparing events of different
known origins. 
Events due to external $\gamma$-radioactivity
are located at shallow depth. For instance,
$\gamma$-rays of 60~keV and 662~keV from 
$^{241}$Am and $^{137}$Cs sources 
would deposit energy in Ge mostly at depth
characterized by the attenuation length of 0.96~mm and
27~mm, respectively.
On the other hand,
events due to internal radioactivity, like X-rays
at 10.37~$\keVee$ from $^{68,71}$Ge K-shell
electron capture to become Ga,
are uniformly distributed in the bulk fiducial volume.
TA $\tau$-distributions of nPCGe and pPCGe
for these events are displayed in 
Figures~\ref{fig::taudist}a\&b, respectively. 
Events with AC$^+$$\otimes$CR$^+$ tags, useful
for {\it in situ} calibration purpose, are also included.

The Bulk-to-Surface event ratios of various sources
are different in pPCGe, verifying the expectation that
these events have different spatial distributions
in the detectors.
On the other hand, 
TA $\tau$-distributions of Bulk samples
are identical among the different samples, 
demonstrating that detector response is uniform over 
the entire fiducial volume.
Accordingly, it is justified to adopt {\it in situ}
physics background events
with AC$^+$$\otimes$CR$^+$ (plus ``Bulk'' tag for pPCGe)
to derive the efficiencies
of neutrino- and WIMP-induced signals in
AC$^-$$\otimes$CR$^-$ samples,
where the feature of 
single-site events uniformly distributed
within the detector volume is identical to that 
of the internal Ga X-rays.

\item[(iii)]{\bf Background Understanding}

An interesting note is that 
AC$^+$$\otimes$CR$^+$ samples of pPCGe
shown in Figure~\ref{fig::taudist}b have their
B-to-S ratio and $\tau$-distributions 
of S-events similar
to those of $^{137}$Cs but not $^{241}$Am. 
This observation
indicates that the AC$^+$$\otimes$CR$^+$ events
at this specified energy range (8$-$12~$\keVee$) 
are mostly due to external $\gamma$-rays of 
MeV-range energy (rather than 10-keV range)
induced by cosmic-rays traversing in the 
vicinity of the detector.
It follows that the studies on timing profiles of 
S-background events may reveal their origins. 
This is one of the subjects of our on-going research.

\end{description}


\subsubsection{Combined Background Spectra}

Data were taken with various Ge detectors at KSNL.
Background events are identified and suppressed
when they are
(i) correlated with the AC and/or CR systems,
(ii) induced by electronic noise with anomalous timing structures, or
(iii) located at the surface of pPCGe.
The various efficiency factors discussed in this article,
namely $\trigeff , \daqeff , \CReff , \ACeff , \bfeff$,
$\effbs$ and $\lmbdbs$, and their uncertainties
are applied.
Their typical values are summarized in 
Table~\ref{tab::summary-selection} and 
stated in the respective figures.
The combined efficiency for nPCGe 
is 91\% above its noise-edge of 373~$\eVee$, while
that for pPCGe is depicted in 
Figure~\ref{fig::combinedeff},
showing the values before and after 
BS-selection.

The electronic noise-edge defines 
the lower reach of the extraction of physics signals
for AC$^-$$\otimes$CR$^-$ events which
are uncorrelated with other detector systems.
The background level at threshold is of
$\mathcal{O}(10 ~ \cpkkd)$ at KSNL.
The evolution of the energy spectra
from raw data via various selection procedures 
is displayed in Figures~\ref{fig::pcspect}a\&b 
for nPCGe and pPCGe with 54.6~kg-days and 46.2~kg-days of data, 
respectively.
X-ray lines from internal radioactivity become 
visible after these selections, and are adopted for
energy calibration.

The electronic noise-edge of 
AC$^{-}$$\otimes$CR$^{-}$ events 
in pPCGe is 311~$\eVee$, as shown in
Figures~\ref{fig::spectra}b\&\ref{fig::pcspect}b.
This threshold applies also to the spectra
following the subsequent BS-cut and  
($\effbs , \lmbdbs$)-correction. 

As a quantitative illustration, the 
BS-correction applied to 
the threshold bin of 300-350~$\eVee$
would modify a raw rate of
$\rm{B = 98 \pm 20}$ 
in $\cpkkd$ unit.
The efficiency factors 
$( \effbs , \lmbdbs ) = ( 0.80 \pm 0.07 , 0.64 \pm 0.01 )$
contribute a boost factor of 
$ [ 1/( \effbs + \lmbdbs - 1 ) ] =  2.25 $ to the
statistical uncertainty, giving
a corrected value of $\rm{B_0 = 36 \pm 17}$.
The systematic uncertainty is $\pm 19$, 
estimated by variations of parameter choice
in the derivation of ($\effbs , \lmbdbs$)~\cite{bsel2014}.

The anomalous surface effects in nPCGe are negligible,
so that corrections with ($\effbs , \lmbdbs$) 
are not applicable.
Physics events can be extracted down to
the noise-edge of 373~$\eVee$,
as shown in Figures~\ref{fig::spectra}c\&\ref{fig::pcspect}a.
The noise-edges in both cases,
as indicated in Figures~\ref{fig::response}b\&c,
are above the range where pulser events exhibit non-linear 
and anomalous response. 

The  AC$^+$$\otimes$CR$^+$ spectra for nPCGe and pPCGe 
are depicted in the insets of 
Figures~\ref{fig::pcspect}a\&b, respectively.  
The triple coincidence of Ge with the AC and CR detector 
systems with AC$^+$$\otimes$CR$^+$ tags
allow physics signals
to be extracted to below the electronic noise-edge
defined by anti-coincident AC$^-$$\otimes$CR$^-$ samples.
These events can serve as reference samples for efficiency
measurements of neutrino- and WIMP-induced events.
The energy threshold values are reduced 
from 373~$\eVee$ to 237~$\eVee$ and
from 311~$\eVee$ to 197~$\eVee$ for
nPCGe and pPCGe, respectively.
Below these energies and above the trigger level, 
electronic self-trigger noise events still overwhelm, despite
having coincidence tags with two detector systems.



\section{Summary and Prospects}

Germanium detectors with sub-keV sensitivities have
opened windows for studies on SM and exotic
neutrino interactions as well as for searches of light
WIMPs. This article documents our efforts on the 
characterization and optimization of the detector response in
energy domains near electronic noise-edge,
where the signal amplitude is comparable to 
that of pedestal fluctuations.

Both pPCGe and nPCGe represent novel
advances in Ge detector techniques to measure events
with sub-keV energy depositions, and the results of 
Figures~\ref{fig::pcspect}a\&b and Table~\ref{tab::summary-selection}
represent improved performance among this class of detectors
over those from previous efforts listed in Table~\ref{tab::survey}, 
lowering the detector
threshold values from $\sim 500 ~ \eVee$
to $\sim 300 ~ \eVee$.

The potential scientific reach depends 
on the achievable detector threshold.
Ongoing R\&D efforts are pursued with this goal,
via optimizations of hardware configurations, 
JFET and ASIC electronic components, as well
as software pulse-shape discrimination techniques
below electronic noise-edge.
We note in particular the novel idea of 
internal amplification
in Ge ionization detectors~\cite{geintamp}.

A new development is the demonstration of 
the ``bolometric amplification'' concept
in the CDMSlite experiment~\cite{cdmslite}
with Ge crystals operating at
cryogenic temperatures. Research programs are pursued
to perfect the technique which offers the potential
of bringing detector threshold 
to $\mathcal{O}( 10~ \eVee )$.
Higher cost and technical complications associated with 
operation at sub-Kelvin temperatures, however, may limit its
range of applications.

\section{Acknowledgments}

This work is supported by
the Academia Sinica Investigator Award 2011-15,
contracts 99-2112-M-001-017-MY3, 
102-2112-M-001-018, 
103-2112-M-001-024 and 104-2112-M-001-038-MY3
from the Ministry of Science and Technology, Taiwan.
HTW is grateful for  
a grant from the Simons Foundation 
and for the hospitality of the Aspen Center for Physics,
where part of the manuscript was composed. 
The authors are grateful to the efforts 
of Ms. J. Wu and Prof. N. Akchurin 
on text editing and proof-reading.

\end{document}

%% file: author_nedge.tex
\newcommand{\as}{Institute of Physics, Academia Sinica,
Taipei 11529, Taiwan.}
\newcommand{\thu}{Department of Engineering Physics, Tsinghua University,
Beijing 100084, China.}
\newcommand{\metu}{Department of Physics,
Middle East Technical University, Ankara 06531, Turkey.}
\newcommand{\deu}{Department of Physics,
Dokuz Eyl\"{u}l University, Buca, \.{I}zmir 35160, Turkey.}
\newcommand{\ciae}{Department of Nuclear Physics,
Institute of Atomic Energy, Beijing 102413, China.}
\newcommand{\bhu}{Department of Physics, Institute of Science,
Banaras Hindu University, Varanasi 221005, India.}
\newcommand{\nku}{Department of Physics, Nankai University,
Tianjin 300071, China.}
\newcommand{\scu}{Department of Physics, Sichuan University,
Chengdu 610065, China.}
\newcommand{\ks}{Kuo-Sheng Nuclear Power Station,
Taiwan Power Company, Kuo-Sheng 207, Taiwan.}
\newcommand{\canberra}{
CANBERRA-Lingolsheim Facility, Lingolsheim 67380, France.
}
\newcommand{\corr}{htwong@phys.sinica.edu.tw}
\newcommand{\corrls}{lakhwinder@gate.sinica.edu.tw}
\newcommand{\corrms}{manoj@gate.sinica.edu.tw}
\newcommand{\aksadd}{
Department of Physics, University of South Dakota, Vermillion,
South Dakota 57069, USA.}
\newcommand{\gkkadd}{Physics Department, KL University, 
Guntur 522502, India.}

\author{ A.K.~Soma }  \altaffiliation[Present Address: ]{ \aksadd } \affiliation{ \as } \affiliation{ \bhu }
\author{ M.K.~Singh } \altaffiliation[Corresponding Author: ]{ \corrms }  \affiliation{ \as } \affiliation{ \bhu }
\author{ L.~Singh }  \altaffiliation[Corresponding Author: ]{ \corrls } \affiliation{ \as } \affiliation{ \bhu }
\author{ G.~Kiran Kumar } \altaffiliation[Present Address: ]{ \gkkadd } \affiliation{ \as }
\author{ F.K.~Lin }  \affiliation{ \as }
\author{ Q.~Du }  \affiliation{ \scu } 
\author{ H.~Jiang }  \affiliation{ \thu } 
\author{ S.K.~Liu } \affiliation{ \scu } \affiliation{ \thu }
\author{ J.L.~Ma }  \affiliation{ \thu } 
\author{ V.~Sharma }  \affiliation{ \as } \affiliation{ \bhu }
\author{ L.~Wang } \affiliation{ \thu }
\author{ Y.C.~Wu } \affiliation{ \thu }
\author{ L.T.~Yang }  \affiliation{ \thu } 
\author{ W.~Zhao }  \affiliation{ \thu } 
\author{ M.~Agartioglu }  \affiliation{ \as } \affiliation{ \deu }
\author{ G.~Asryan }  \affiliation{ \as }
\author{ Y.Y.~Chang }  \affiliation{ \as }
\author{ J.H.~Chen }  \affiliation{ \as }
\author{ Y.C.~Chuang }  \affiliation{ \as }
\author{ M.~Deniz } \affiliation{ \as } \affiliation{ \deu } \affiliation{ \metu }
\author{ C.L.~Hsu }  \affiliation{ \as }
\author{ Y.H.~Hsu }  \affiliation{ \as }
\author{ T.R.~Huang }  \affiliation{ \as }
\author{ L.P.~Jia }  \affiliation{ \thu } 
\author{ S.~Kerman }  \affiliation{ \as } \affiliation{ \deu }
\author{ H.B.~Li }  \affiliation{ \as }
\author{ J.~Li }  \affiliation{ \thu } 
\author{ F.T.~Liao }  \affiliation{ \as }
\author{ H.Y.~Liao }  \affiliation{ \as }
\author{ C.W.~Lin }  \affiliation{ \as }
\author{ S.T.~Lin }  \affiliation{ \as } \affiliation{ \scu }
\author{ V.~Marian }  \affiliation{ \canberra }
\author{ X.C.~Ruan } \affiliation{ \ciae }
\author{ B.~Sevda }  \affiliation{ \as } \affiliation{ \deu }
\author{ Y.T.~Shen }  \affiliation{ \as }
\author{ M.K.~Singh } \affiliation{ \as } \affiliation{ \bhu }
\author{ V.~Singh }  \affiliation{ \bhu }
\author{ A.~Sonay }  \affiliation{ \as } \affiliation{ \deu }
\author{ J.~Su }  \affiliation{ \thu } 
\author{ V.S.~Subrahmanyam }  \affiliation{ \as } \affiliation{ \bhu }
\author{ C.H.~Tseng }  \affiliation{ \as }
\author{ J.J.~Wang }  \affiliation{ \as }
\author{ H.T.~Wong } \altaffiliation[Corresponding Author: ]{ \corr } \affiliation{ \as }
\author{ Y.~Xu } \affiliation{ \as } \affiliation{ \nku }
\author{ S.W.~Yang }  \affiliation{ \as }
\author{ C.X.~Yu } \affiliation{ \as } \affiliation{ \nku }
\author{ Q.~Yue } \affiliation{ \thu }
\author{ M.~Zeyrek } \affiliation{ \metu }

%% file: table_summary_performance.tex

\begin{table*}
\caption{
Summary table of performance parameters 
of Ge detectors in this study.
The pulse maxima ($\eamp$) is adopted as energy estimator.
}
\begin{ruledtabular}
\begin{tabular}{lccccc}
Performance Parameters & CoaxGe & ULEGe & pPCGe & nPCGe & Uncertainties  \\
 & &  & &  & (\%) $^\amalg$ \\ \hline
\hspace*{0.1cm} Modular Mass (g) &1000 & 5 & 500 & 500 & $-$ \\
\hspace*{0.1cm} RESET Amplitude (V)  & N/A $^\dagger$ & 8.0 &6.8 & 6.8  &  $-$ \\
\hspace*{0.1cm} RESET Time Interval (ms) & N/A $^\dagger$ & 
$\sim$700 & $\sim$160 & $\sim$170  &  $-$ \\
\hspace*{0.1cm} Pedestal Noise  & & & &  & \\
\hspace*{0.6cm} Pedestal Profile RMS $\amprms$ ($\eVee$) & 812 & 33 & 41 & 49  & 2.6 \\
\hspace*{0.6cm} Area RMS $\qrms$ ($\eVee$) & 840 & 30 & 58 & 52  &  3.1 \\
\hspace*{0.1cm} Pulser Width & & & &  & \\
\hspace*{0.6cm} FWHM ($\eVee$) & 1566 & 87 & 110 & 122 & 1.5 \\
\hspace*{0.6cm} RMS ($\eVee$) & 665 & 37 & 47  & 52 & 1.5 \\
\hspace*{0.1cm} X-Ray Line Width & Ga-K & $^{55}$Fe 
& Ga-K &  Ga-K & $-$ \\
\hspace*{0.6cm} RMS ($\eVee$) & 880 & 64 & 87 & 104 & 3.4 \\
\hspace*{0.1cm} Electronic Noise-Edge for Raw Spectra ($\eVee$) & 
4900 &  220 & 228 &  285 & 1.8 
\end{tabular}
\end{ruledtabular}
\begin{flushleft}
$^\amalg$ Uncertainties of pPCGe measurements are quoted as illustration. 
Other detectors have similar levels.\\
$^\dagger$ Resistance feedback preamplifier is used in CoaxGe,
so that the RESET timing structures are not-applicable (N/A).\\
\end{flushleft}
\label{tab::summary-performance}
\end{table*}

%% file: table_survey.tex

\begin{table*}
\caption{
Summary table of published performance parameters from
previous studies on Ge detectors in which
low energy threshold plays crucial roles.
}
\begin{ruledtabular}
\begin{tabular}{lccccc}
Studies & Luke et al. &
IGEX & CoGeNT & MALBEK &  CDEX-1A \\
& \cite{lukepcge}  & \cite{igex} & \cite{cogent} &
\cite{malbek} & \cite{cdex1} \\ \hline
\hspace*{0.1cm} Modular Mass (g) & 800 & 2200  & 440  & 465 & 994 \\
\hspace*{0.1cm} Pedestal Profile RMS $\amprms$ ($\eVee$) 
& 115  &  $-$ & 69  &  70  & 51  \\
\hspace*{0.1cm} X-Ray Line & $-$ & 
 Pb-X  & Ga-K &  Ga-K  & Ga-K \\
\hspace*{0.6cm} Energy ($\keVee$) &  $-$ & 75 & 10.4 & 10.4 & 10.4 \\
\hspace*{0.6cm} Resolution RMS ($\eVee$) &  $-$ & 340 & 111 & 117 & 91 \\
\hspace*{0.1cm} Detector Threshold ($\eVee$) & 700  & 4000    
& 400  & 600 & 400 \\ 
\end{tabular}
\end{ruledtabular}
\label{tab::survey}
\end{table*}

%% file: table_summary_selection.tex

\begin{table*}
\caption{
Summary table of signal selection procedures
in Ge detectors.
The pulse maxima ($\eamp$) is adopted as energy estimator.
Low-background data taken at KSNL are used.
}
\begin{ruledtabular}
\begin{tabular}{lccccc}
Signal Selection & CoaxGe & ULEGe & pPCGe & nPCGe & Uncertainties  \\
  & &  & &  & (\%) $^\amalg$ \\ \hline
\hspace*{0.1cm} 1) Trigger & & & & & \\
\hspace*{0.6cm} Pedestal Profile RMS $\amprms$ ($\eVee$) 
& 812 & 33 & 41 & 49  & 2.6 \\
\hspace*{0.6cm} Selected Trigger Level $\Delta$ ($\amprms$)  
& 4.3 & 4.3 & 4.2 & 4.2 & $-$ \\
\hspace*{0.6cm} Trigger Threshold at $\trigeff = 50\%$ ($\eVee$) 
& 3500 &  142 & 171 & 204 & 1.4 \\
\hspace*{0.1cm} 2) DAQ & & & & & \\
\hspace*{0.6cm} $\daqeff$ (\%) & 
N/A $^\ast$ & N/A  $^\ast$ & 86 & 87 & $< 0.1$ \\
\hspace*{0.1cm} 3) Analysis Selection & & & & & \\
\hspace*{0.6cm} a) Other Detector Systems & & & & & \\
\multicolumn{1}{l}{\hspace*{1.0cm} i) $\CReff$ 
of Cosmic-Ray Vetos  $-$} &
\multicolumn{5}{c}{Typical Range \hspace*{0.5cm} $ 92.1 \pm 0.02$ ~ \%} \\
\multicolumn{1}{l}{\hspace*{1.0cm} ii) $\ACeff$
of Anti-Compton NaI(Tl) $-$} &
\multicolumn{5}{c}{Typical Range \hspace*{0.5cm} $ 99.5 \pm 0.06$ ~ \%} \\
\multicolumn{6}{l}{\hspace*{0.6cm} b) $\epsilon$ of Basic 
Filters at Noise-Edge (\%)} \\
\hspace*{1.0cm} i) $\Delta t^+$-$\Delta t^-$ (Figure~\ref{fig::basicfilter}a) 
& N/A $^\dagger$ &  N/A $^\ddagger$ & 97.9 & N/A $^\ddagger$ & $< 0.1$ \\
\hspace*{1.0cm} ii) Pedestal Range (Figure~\ref{fig::basicfilter}b) 
& N/A $^\ddagger$  & N/A $^\ddagger$ & 99.7 & 99.8 & $< 0.1$ \\
\hspace*{1.0cm} iii) Pulse Maximal Position (Figure~\ref{fig::basicfilter}c) 
& N/A $^\ddagger$ & N/A $^\ddagger$ & 99.97 & 99.98 & $< 0.1$ \\
\hspace*{0.1cm} 4) Physics Events Threshold ($\eVee$) & & & & & \\
\hspace*{0.6cm} a) AC$^{-}$$\otimes$CR$^{-}$ & 5600 & 230 & 311 & 373 & 1.3 \\
\hspace*{0.6cm} b) AC$^{-}$$\otimes$CR$^{-}$$\otimes$Bulk 
& N/A$^\ddagger$ & $--$ & 311 & $--$ & 2.8 \\
\hspace*{0.6cm} c) AC$^{+}$$\otimes$CR$^{+}$
& N/A$^\ddagger$ &  N/A$^\ddagger$ &  197  & 237  & 2.3  \\ 
\end{tabular}
\end{ruledtabular}
\begin{flushleft}
$^\amalg$ Uncertainties of pPCGe measurements are quoted as illustration.
Other detectors have similar levels.\\
$^\ast$ DAQ were shared with a different 
measurement~\cite{texononuecsi}, such that 
the efficiencies are not relevant to this discussion.\\
$^\ddagger$ Hardware configurations and analysis procedures 
are different in these early measurements.
\end{flushleft}
\label{tab::summary-selection}
\end{table*}